\documentclass[11pt, preprint, floatfix, aps, prl]{revtex4-1}

\usepackage{natbib}
\usepackage{graphicx}
\usepackage{amsmath}
\usepackage{fullpage}
\usepackage{hyperref}
\usepackage{bm}

\hypersetup{
    colorlinks=true,       
    linkcolor=black, 
    citecolor=blue, 
    filecolor=magenta,      
    urlcolor=cyan           
}

\begin{document}

\title{A dark incompressible dipolar liquid of excitons}

\author{Kobi Cohen}
\author{Yehiel Shilo}
\author{Ronen Rapaport}
\affiliation{Racah Institute of Physics, The Hebrew University of Jerusalem, Jerusalem 91904, Israel}
\author{Ken West}
\author{Loren Pfeiffer}
\affiliation{Department of Electrical Engineering, Princeton University, Princeton, New Jersey 08544, USA}

\begin{abstract}

The possible phases and the nano-scale particle correlations of two-dimensional interacting dipolar particles is a long-sought problem in many-body physics. Here we observe a spontaneous condensation of trapped two-dimensional dipolar excitons with internal spin degrees of freedom from an interacting gas into a high density, closely packed liquid state made mostly of dark dipoles. Another phase transition, into a bright, highly repulsive plasma is observed at even higher excitation powers. The dark liquid state is formed below a critical temperature $T_c\approx 4.8K$, and it is manifested by a clear spontaneous spatial condensation to a smaller and denser cloud, suggesting an attractive part to the interaction which goes beyond the purely repulsive dipole-dipole forces. Contributions from quantum mechanical fluctuations are expected to be significant in this strongly correlated, long living dark liquid. This is a new example of a two-dimensional atomic-like interacting dipolar quantum liquid, but where the coupling of light to its internal spin degrees of freedom plays a crucial role in the dynamical formation and the nature of resulting ground state.
\end{abstract}

\maketitle

Cold two-dimensional dipolar gases are an intriguing and theoretically challenging example of many-body interacting quantum fluids. Theoretical attempts to understand the phases and correlations of such dipolar fluids revealed a very rich behavior, with predictions of several exotic quantum and classical phases and correlation regimes \cite{santos_bose-einstein_2000, buchler_strongly_2007, lahaye_physics_2009, astrakharchik_quantum_2007, laikhtman_correlations_2009, laikhtman_exciton_2009, moroni_coexistence_2014} that are absent in weakly interacting Bose gases. These predictions sparked intensive experimental efforts in both atomic and molecular systems \cite{lu_strongly_2011, gaj_molecular_2014, firstenberg_attractive_2013, jin_polar_2011, yan_observation_2013}, as well as in two dimensional (2D) dipolar excitons in  semiconductor bilayer systems (also known as indirect excitons - IXs) \cite{eisenstein_bose-einstein_2004,high_spontaneous_2012,shilo_particle_2013,kazimierczuk_giant_2014}.
These IXs are composed of a coulomb-bounded, yet spatially separated electron-hole pairs in an electrically biased semiconductor double quantum well (DQW) structure \cite{butov_condensation_2004, rapaport_experimental_2007}.
Recent theoretical analysis of the correlations in such a dipolar IX system predicts that in a very large range of densities the onset of quantum degeneracy is accompanied by the set-in of multi-particle interactions and correlations, so that dipolar IXs should behave as a strongly-interacting liquid rather than as a weakly interacting Bose-Einstein condensate (BEC) \cite{laikhtman_correlations_2009, laikhtman_exciton_2009}. What makes this IX system even more unique is the additional internal spin degrees of freedom of excitons in GaAs-based DQWs, allowing to explore the effect of spin on quantum correlations. The IX can be found in four different spin states, namely $S=\pm 1,\pm 2$. The IXs with $S=\pm 1$ are optically active and are named "bright" while the ones with $S=\pm 2$ are optically inactive and are named "dark" \cite{poem_accessing_2010}. This dark IX thus have a much longer lifetime.  Due to spin-dependent exchange interaction between electrons and holes, the dark IXs are slightly lower in energy than the bright IXs \cite{maialle_exciton_1993}, and therefore it was predicted that the ground state of the dipolar exciton fluid is dark and that a macroscopic phase transition to a dark BEC should be observed \cite{combescot_bose-einstein_2007}. 

From the theoretical point of view, it seems that the picture is still incomplete, as no theory takes into account both dipolar interactions and the internal spin degrees of freedom to predict the origin and nature of the ground state of the IX fluid. On the experimental side, there are reports on the onset of coherent phenomena expected from a quantum degenerate dilute gas of IXs \cite{high_spontaneous_2012,high_condensation_2012,high_spin_2013,alloing_evidence_2014}, but other works show experimental evidence for a formation of a highly correlated interacting fluid \cite{shilo_particle_2013}  and for an exciton liquid \cite{stern_exciton_2014} and of a notable spontaneous darkening of a correlated dipolar exciton fluid below a certain temperature \cite{shilo_particle_2013}.  It is therefore an open challenge to fully and unambiguously determine the formation mechanism and properties of the low temperature highly correlated collective dipolar state. Can the combination of the strong dipolar interactions, quantum degeneracy, together with the very long lifetime of the dark particles lead to a long lived, \textit{dark quantum dipolar liquid}?  

In this paper we report on the first observation of a spontaneous gas-liquid-plasma transitions of a trapped IX cloud at low temperatures, and show that the liquid phase is made up of mostly condensed dark particles, that it has the characteristic parameter range of an incompressible closely-packed quantum liquid, and that the underlying physics probably does not result from purely dipolar forces, but from more complex interactions at high densities, which have to do with the short range interactions due to the internal spin degrees of freedom.

\begin{figure}[ht!]
  \centering
  \includegraphics[width=0.45\textwidth]{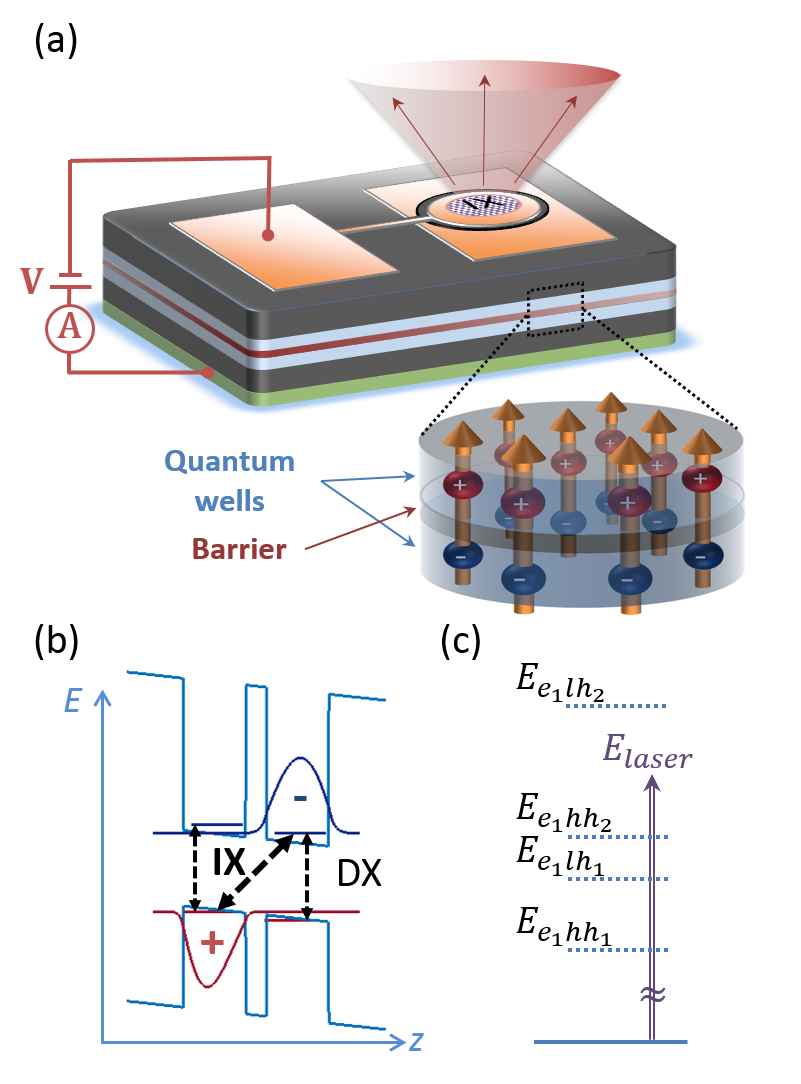}
  \caption{(a) A sketch of the sample structure, showing two quantum wells (light blue) separated by a thin barrier (red). An electric field is applied perpendicular to the growth axis between a transparent $10nm\:Ti$ top circular gate and $n^+$ doped GaAs layer at the bottom. A non-resonant optical excitation using a CW diode laser creates a trapped IX fluid under the gate, as illustrated. The tunneling current through the sample is also monitored. (b) Schematics of a calculated energy bands diagram for the structure, showing the conduction and valence bands and the electron and hole wave functions, respectively. IX marks the indirect exciton transition, whereas DX marks the direct exciton. (c) The first energy levels of direct exciton transitions (dashed lines), of either electrons $(e)$ and heavy holes $(hh)$ or light holes $(lh)$ are marked, as well as the laser excitation at a wavelength of $780nm$ (solid arrow), illustrating the non resonant excitation of the quantum wells.} \label{fig:fig1}
\end{figure}

\begin{figure*}[ht!]
  \centering
  \includegraphics[width=0.8\textwidth]{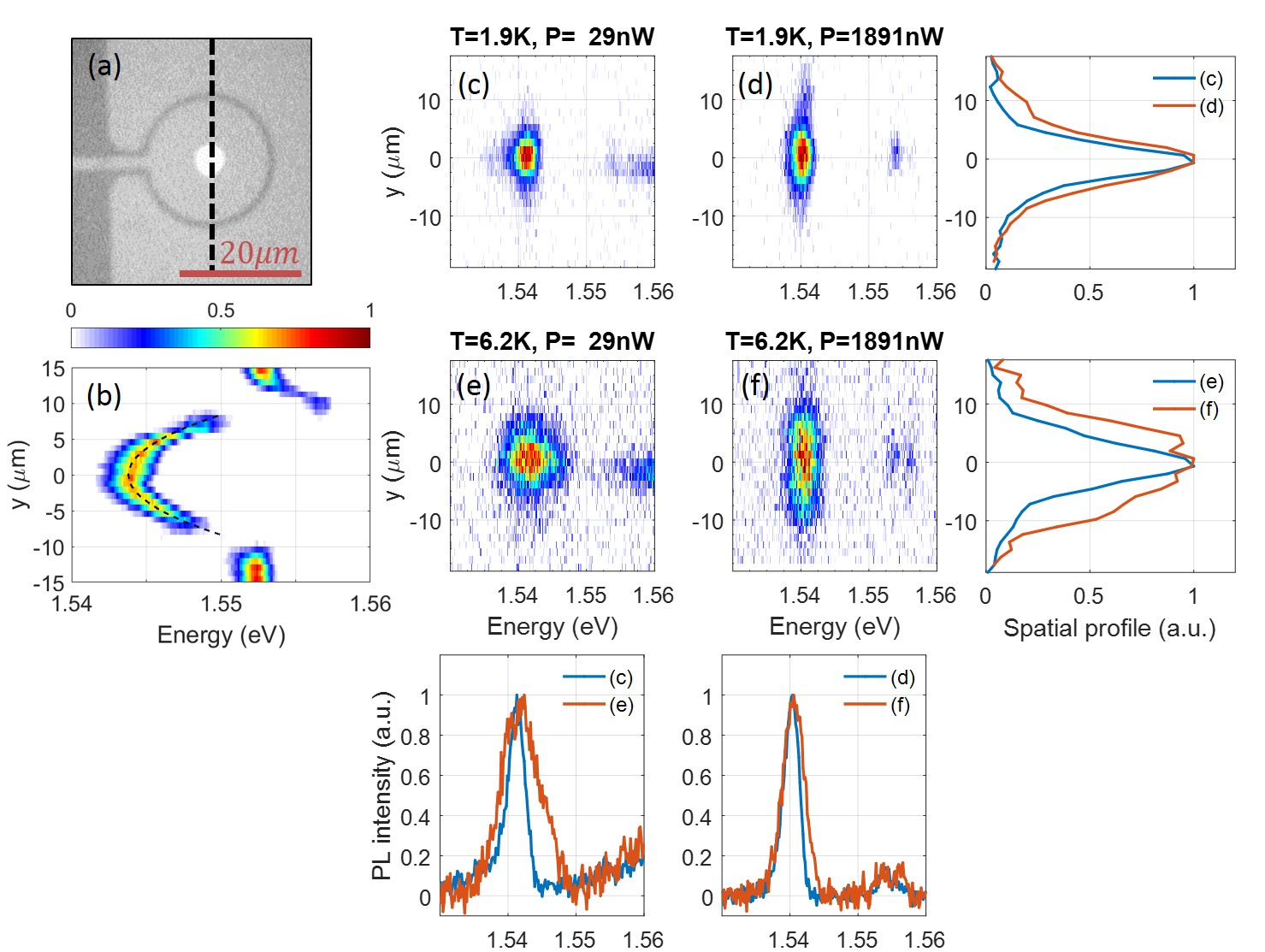}
  \caption{(a) A microscope image of a $20\mu m$ diameter electrostatic trap (Xtrap) surrounded by a guard gate. The bright spot at the center is the exciting laser image. (b) Spatial-spectral profile of the Xtrap, obtained by scanning a weak and focused excitation point along the line marked in (a). (c-f) Spatial-spectral PL intensity for temperatures $T=1.9, 6.2K$ and excitation powers $P=29, 1891nW$, all taken on a CEL by adjusting the voltage. The horizontal axis shows the PL energy, and the vertical axis is a cross-section of the position along the Xtrap. The peak with the lower energy ($E_{IX}\approx 1.541eV$) is PL from the indirect exciton (IX), and the weak emission with the higher energy ($E_{DX}\approx 1.554eV$) is PL from the direct exciton (DX). Right: spatial intensity profiles of (c-f), obtained by integration along the spectral axis, showing (c,d) very minor spatial expansion of the PL for increasing $P$ at low $T$, in contrast to (e,f) a substantial increase of the PL area with excitation power at high $T$. Bottom: the extracted spectra from (c-f), obtained by integration along the spatial axis, illustrating (c,e) a narrower line-shape at low $T$ and, (d,f) a reduction of the linewidth to a symmetric and narrow emission line-shape as the excitation power increases.}
  \label{fig:fig2}
\end{figure*}

To have a well defined and highly controlled system, we trap a cloud of IXs, optically excited in a GaAs based DQW structure under an electrostatic trap gate \cite{rapaport_electrostatic_2005}, as is depicted schematically in Fig.\ref{fig:fig1}. The IXs are excited non-resonantly at the center of this quasi-parabolic electrostatic trap (denoted Xtrap), and its photoluminescence (PL) is measured under steady-state conditions (see Fig.\ref{fig:fig2}(a,b) for the Xtrap geometry and its energy profile respectively).

The spatial-spectral emission profile of the IXs is imaged on a CCD, allowing us to follow the size and shape of the IX cloud. Furthermore, we use a constant energy line method (CEL, see \cite{shilo_optical_2014} and SI) in which we can keep the energetic and electric environment of the exciton fluid fixed. This is achieved by adjusting the magnitude of the applied electric field $F$ on the DQW to keep the band tilting and thus the difference between the direct exciton (DX) and the IX transition energies, $E_{DX}-E_{IX}$ fixed while varying the optical excitation power and temperature. With this method we can extract information on the relations between the energy resulting from the many-body interactions between the IXs - $\Delta E$ \cite{shilo_particle_2013} and their density, while fixing the \emph{single-particle} properties of an IX (either bright or dark). This is because these single IX properties, such as its radiative and non-radiative lifetimes and its effective dipole size $d$ are determined by the same band tilting which also uniquely determines $E_{DX}-E_{IX}$.

Four exemplary spatially resolved PL (along a central cross-section of the circular Xtrap) are shown in Fig.\ref{fig:fig2}(c-f) for non-resonant excitation at the center of the Xtrap. These measurements show two different temperatures and excitation powers, while the applied voltage was adjusted in each of the measurements to maintain a constant energy ($E_{DX}-E_{IX} \approx 13meV$) for all of them. Two excitonic PL lines are observed for all the spectra, the high energy line (very weak) corresponds to the DX while the low energy line correspond to the IX. 
The PL is then spectrally integrated to obtain spatial profiles of the emission (Fig.\ref{fig:fig2} right) and spatially integrated to obtain a single spectrum for each measurement (Fig.\ref{fig:fig2} bottom). Such spectra are obtained and analyzed for different experimental conditions as is described next.

\begin{figure}[ht!]
\centering
\includegraphics[width=0.45\textwidth]{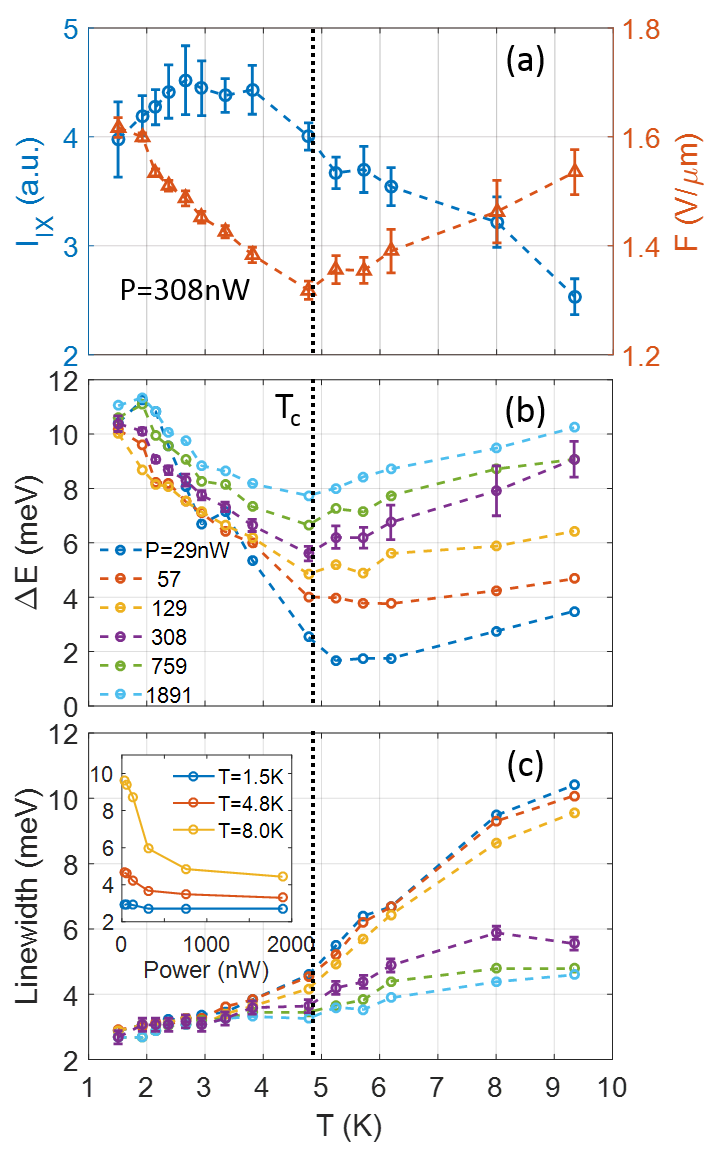}
\caption{Results obtained for a CEL with $E_{IX}=1.541eV$. (a) Total intensity of the IX line - $I_{IX}$ (left axis, blue circles) and the electric field $F$ (right axis, orange triangles) needed in order to maintain this CEL, as a function of temperature for a fixed excitation power of $P\approx308 nW$. (b) The blue-shift energy $\Delta E$ and (c) the IX PL linewidth as a function of temperature $T$ for several different excitation powers. In the high temperature regime ($T>T_c$), $\Delta E$ decreases with temperature, indicating increased particle correlations. For $T<T_c$, $\Delta E$ \textit{increases} as the temperature is lowered and saturates to an almost fixed value, regardless of the excitation power. This marks the condensation of IXs into a fixed, high density liquid state. The reduction of the PL linewidth with decreasing $T$ to a small fixed value points at reduced density fluctuations, as expected from a liquid phase. The inset in (c) shows the dependence of the PL linewidth on excitation power for three different temperatures. 
}
\label{fig:fig3}
\end{figure}

Figure \ref{fig:fig3}(a) shows the total intensity $I_{IX}(T)$ extracted from the area under the spectral line of the IXs, obtained at different temperatures for a fixed excitation power ($P\approx 308nW$) on a CEL with a fixed $E_{IX}\approx 1.541eV$. Also shown is the applied electric field $F$ that was required in order to maintain the CEL at the aforementioned energy.
By performing such CEL experiments for excitation powers $P$ spanning almost two orders of magnitude for each $T$, we extract the many-body interaction term $\Delta E(T,P)$ in a straightforward way: $\Delta E(T,P)=eF(T,P)d-eF(T,0)d$ where $F(T,P)$ is the applied field required for a fixed $E_{DX}-E_{IX}$ at a given $(T,P)$, and $d$ is the effective distance between the centers of the two QWs (see SI for more detailed information on this method). These $\Delta E$ values and the spectral linewidth of the IXs PL are plotted as a function of $T$ for different $P$ values in Fig.\ref{fig:fig3}(b,c).

In the higher temperature range ($T>4.8K)$, $\Delta E$ shown in Fig.\ref{fig:fig3}(b) decreases with decreasing temperature but increases with increasing excitation power. The increase of $\Delta E$ with $P$ is not surprising:  the magnitude of the repulsive dipole-dipole interactions increase with increasing total density $n$, i.e., $\Delta E=C(n,T) \cdot n$ (where $C$ is the interaction-induced particle correlation function \cite{laikhtman_correlations_2009}), and $n$ increases with increasing $P$. The decrease of $\Delta E$ with decreasing $T$ (for a fixed $P$) at this temperature range was also predicted theoretically \cite{laikhtman_exciton_2009, laikhtman_correlations_2009, zimmermann_excitonexciton_2007} and observed experimentally \cite{shilo_particle_2013}, and it arises from increased interaction-induced particle correlations as the IXs become colder, resulting with a reduction of $C$ and thus of $\Delta E$ . Already at this higher temperature range a reduction of the IX linewidth with increasing $P$ for all temperatures above $4.8K$ is observed, as seen in Fig.\ref{fig:fig3}(c). Such a strong linewidth reduction with increasing particle density indicates that the increase of density reduces the fluctuations of the interaction energy around its average value. This density-induced linewidth narrowing is a strong indication for multi-particle interactions and for deviation from a gas behavior \cite{stern_exciton_2014}.

Surprisingly, moving to the low-$T$ range ($T<4.8K$), $\Delta E$ changes its trend and sharply rises with decreasing $T$, as can be seen in Fig.\ref{fig:fig3}(b). $\Delta E$ slightly "overshoots" to a maximum at $T\approx 2K$ and then slightly decreases and reaches a \textit{constant value}, $\Delta E_l\simeq 10.5 meV$, at the lowest measured temperature, \textit{for all excitation powers} $P$.
The PL linewidth also converge below $T=4.8K$ to a constant low value at the lowest measured temperature, as can be seen in Fig.\ref{fig:fig3}(c). At this temperature, the linewidth also becomes independent of the excitation power, as is seen in the inset.
The sharp increase of $\Delta E$ below $4.8K$ indicates an increase of the total dipolar fluid density $n$ \footnote{This increase cannot be due to a sudden increase in $C$, since particles at lower temperatures have lower kinetic energies, higher spatial correlation, and thus smaller values of $C$}. Furthermore, the fact that $\Delta E\rightarrow\Delta E_l$ at the lowest measured $T$, together with a reduction of the PL linewidth to a small, fixed value suggests that the trapped fluid reaches a fixed density $n_l$ regardless of the excitation power. This observation is consistent with a transition into a condensed liquid state, that has a high density, fixed by the constraints of the physical system.

\begin{figure*}[ht!]
  \centering
\includegraphics[width=0.75\textwidth]{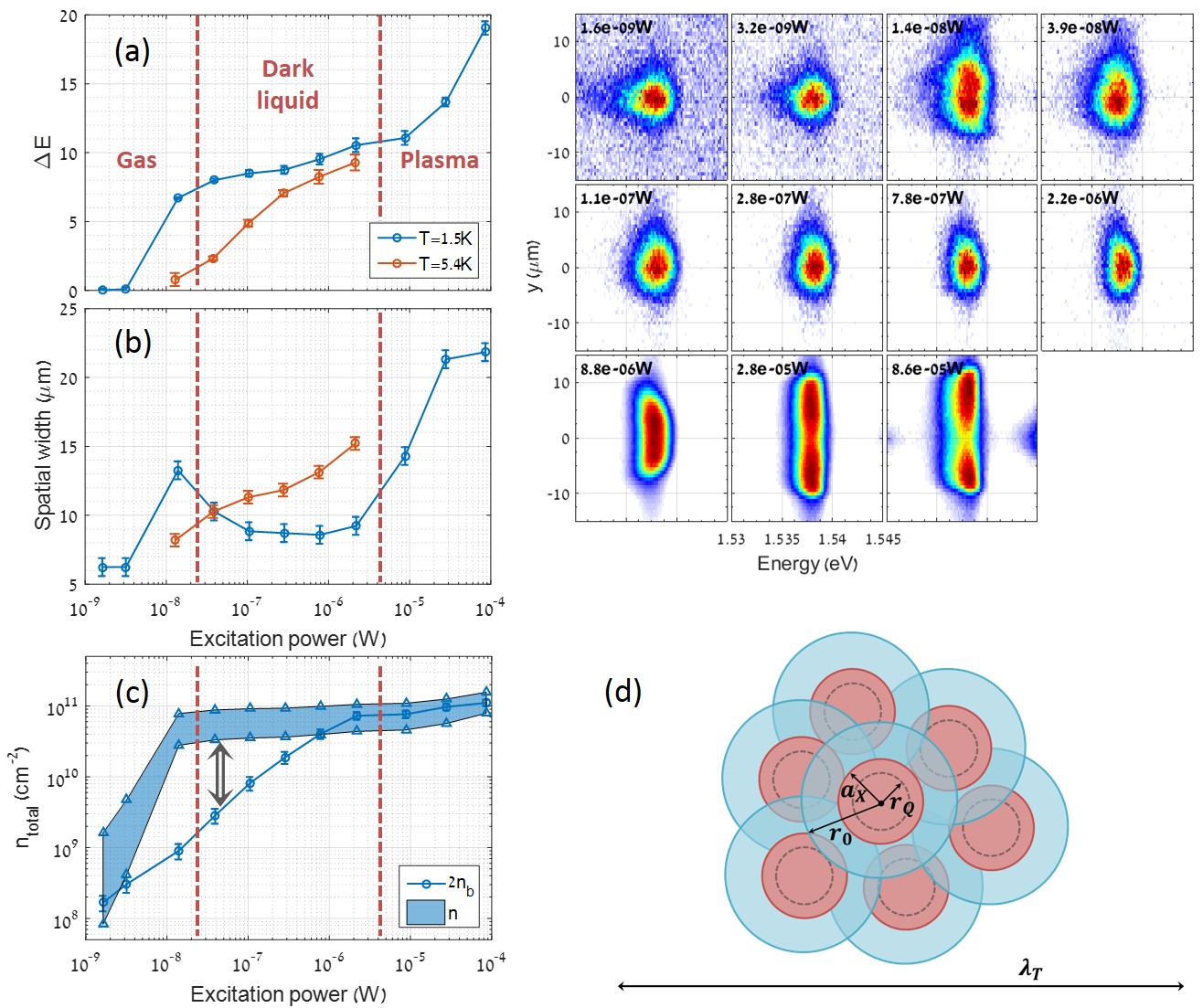}
  \caption{Results obtained for a CEL with $E_{IX}=1.538eV$ with excitation power ranging almost 5 orders of magnitude. Right panels: spatial-spectral photoluminescence images of the trapped IX fluid taken at $T=1.5K$ with increasing excitation power. Left panels: (a) $\Delta E$ and (b) spatial FWHM of the exciton fluid as a function of excitation power at $T=1.5K$ (blue curves), compared with those measured at $T=5.4K$ (red curves). A transition between three different regimes as the excitation power increases is observed only for the low-temperature experiment. (c) Estimated density of the exciton cloud $n_{total}$ as a function of excitation power at $T=1.5K$.
The data obtained from photon-counting (circles) is presented as twice the bright exciton density, $2n_b$ (assuming $n_b = n_d$), together with an expected density range (triangles, shaded region) that is calculated from the measured blueshift energy $\Delta E$ (see text for details). The black vertical arrow indicates the difference $n_{total}-2n_b = n_d-n_b$.
(d) Estimated values of length scales for an exciton liquid: $a_x\approx 15nm$ is the exciton Bohr radius, $r_Q\approx 10nm$ is the quantum mechanical wavefunction radius of an exciton, $r_0\approx 55nm$ is the classical minimal distance between excitons and $\lambda_T\approx 200nm$ is the thermal de-Broglie wavelength (both calculated at $T=2K$). See text and SI for details on the derivation of those values.}
  \label{fig:fig4}
\end{figure*}

Perhaps the most striking evidence for the condensed phase formation and its properties can be seen in Fig.\ref{fig:fig4}. Since in Fig.\ref{fig:fig3} the condensation at low temperatures was observed for all the measured excitation powers, we expanded the excitation power range even further to cover five orders of magnitude, and capture how the phase transitions are driven by the increased rate of optical excitation. The top-right panels of Fig.\ref{fig:fig4} show the spatial-spectral emission profile of the IXs at $T=1.5K<T_c$. Initially, increasing the power results in an expansion of the IX cloud in the trap, as is expected from the mutual repulsive interactions between IXs. However, above a certain threshold power, the cloud \textit{shrinks} with increasing excitation power and reaches a fixed size which almost does not vary over about two orders of magnitude of excitation powers. This marks the condensation to the fixed density liquid phase. When the power is further increased above another threshold value, the cloud starts to expand again in a fast rate. This fast expansion is a signature of a transition to an e-h plasma, driven outwards by the intra-layer e-e and h-h Coulomb repulsion. The spatial profile FWHM and the corresponding $\Delta E$ dependence on the excitation power at $T=1.5K$ are shown by the blue symbols in Fig.\ref{fig:fig4}(a,b) respectively. It can be seen that the initial low power expansion turns into a very strong contraction that is accompanied by a large increase of $\Delta E$, indicating a spontaneous  large increase of density (which we will discuss later on). This is followed by an intermediate power regime (corresponding exactly to the two orders of magnitude power range of Fig. \ref{fig:fig3}), in which both the spatial FWHM and $\Delta E$ are nearly constant, corresponding to an almost fixed density, regardless of excitation power. The high power regime shows a sudden and sharp increase of both the spatial FWHM and of $\Delta E$, consistent with the picture of a strongly driven indirect e-h plasma. The orange symbols show the behavior at $T=5.4K>T_c$, above the condensation temperature. Indeed, non of the features of the condensation are observed, and the IXs expand and increase their density monotonically, consistent with a repulsive dipolar gas phase. 

We can now estimate the saturated density of the condensed phase, $n_l$, from the saturated value $\Delta E_l$. Such an estimate requires a knowledge of the spatial particle correlations and $C$. In Ref. \cite{laikhtman_exciton_2009} a theoretical estimate for $C$ was given for a dipolar exciton liquid. The density estimated from this liquid model yields $n_{l}\approx \left(\frac{\kappa\Delta E}{10e^2d^2}\right)^{2/3} \approx 10^{11}cm^{-2}$, by setting the experimental value $\Delta E_l=10.5meV$ and the dielectric constant $\kappa=13$. A more self-consistent estimate, taking into account the kinetic energy from quantum motion (see SI), gives $n_l\approx 9\cdot10^{10}cm^{-2}$. This density corresponds to an average inter-particle distance $\langle r \rangle=n_{l}^{-1/2}\approx 30nm$. This average inter particle distance is smaller than the effective interaction radius of the dipolar exciton \cite{laikhtman_exciton_2009} $r_0=\left(\frac{e^2d^2}{\kappa k_B T}\right)^{1/3}\approx 55nm$ at $T=2K$, which means that at this density the fluid is deeply inside the multi-particle interaction regime, thus the picture of a dipolar liquid state is self-consistent.

Remarkably, $\langle r \rangle$ is between twice to three times the exciton Bohr radius ($a_X\approx 10-15nm$), namely $\langle r \rangle/a_X\approx 2-3$. Therefore, at this density the IXs are almost closely packed and any further increase in the density significantly increases the overlap of the excitons wave functions. Such an increased overlap should result in a large deviations from the pure dipolar repulsion due to short-range direct and exchange coulomb terms. This specific inter-particle distance might point to the physical mechanism which both drives the spontaneous spatial shrinkage and determines the almost \textit{fixed} density and thus the low compressibility of the condensed liquid. Recent numerical simulations of the two-body interactions of dipolar excitons have suggested that the interaction of two dipolar excitons can become slightly attractive as the inter-particle distances approach $\langle r \rangle\approx 2-3a_X$, and then strongly repulsive scaling as $1/r^6$ at even shorter distances \cite{lee_exciton-exciton_2009}. This inter-particle distance is similar to the estimate of the current experiment.

A natural question now is what can be the mechanism responsible for the observed sharp increase in the IX cloud density as the excitation power increases above the critical power? In principle, a sharp increase in the total density under a fixed excitation power can result from a sharp increase in the \textit{radiative} lifetime of the particles. Noting however that the whole experiment was done under single CEL, where the radiative lifetime of a single IX is kept constant, such an increase is not plausible. The other option is thus that the thermal equality between the populations of dark and bright IXs breaks, and that there is a spontaneous macroscopic accumulation of dark IXs above the critical excitation power. As dark IXs are optically inactive and have much longer lifetimes \cite{shilo_particle_2013}, this should result in a sharp increase of the total density for a fixed excitation power, as is indeed observed in the experiment.  

To check this latter picture, we extracted the total particle density, $n=n_b+n_d$, from the interaction energy $\Delta E$, as plotted in Fig.\ref{fig:fig4}(c). To avoid any model assumption about the relation between $n$ and $\Delta E$ through the unknown correlation parameter $C$, the shaded blue region covers all possible values of $C$, where the lower boundary is given by the uncorrelated gas limit (mean-field) and the upper boundary by the highest correlated limit of a dipolar crystal  \cite{laikhtman_correlations_2009,rapaport_experimental_2007}. The sharp increase of $n$ at the critical power of the gas-liquid phase transition and its saturation up to the second critical power are clear. This behavior of $n$ extracted from $\Delta E$ is then compared with the total density estimated from the bright IX density $n_b$ alone, which we very carefully extracted from the intensity of IX emission (see SI for details). \textit{If we assume} a thermal equality between bright and dark IX populations, i.e., $n_d=n_b$ then $n=2n_b$. 
This total density is plotted by the blue circles in Fig. \ref{fig:fig4}(c). As can be seen, the density of bright IX smoothly increases with excitation power and does not exhibit any sharp increase at the phase transition, and thus it gives a much lower density than the density inferred from $\Delta E$. Since $\Delta E$ inherently depends on the total density $n$, while an assumption of thermal equality of bright and dark IXs population was used to extract $n$ from the PL intensity, it is clear that this assumption strongly \textit{underestimates} the total density all the way to the second critical power where the transition to a plasma is observed, As such a strong underestimate is not observed for the data above $T_c$ (see SI), this strongly supports the picture of condensation of the IXs population into a dark state \cite{combescot_bose-einstein_2007}.
It is interesting to note that the thermal de-Broglie wavelength of excitons at this temperature is estimated to be $\lambda_T=h/\sqrt{2M_Xk_BT}\simeq 200nm$. Therefore, $\langle r \rangle \ll \lambda_T$ so that the IX cloud is quantum degenerate and this should support the condensation of dark IXs \footnote{We note however that this estimate is for a non-interacting Bose gas, while as mentioned in Ref.\cite{laikhtman_exciton_2009} quantum degeneracy could be suppressed down to lower temperatures due to multi-particle correlations}.

Finally, another important length scale that characterizes the liquid phase the is the quantum radius of the IX.  This quantum radius, $r_Q$, is the ground-state wavefunction radius of an exciton confined in the  parabolic-like potential resulting from the interaction with its surrounding excitons in the liquid. With a density $n_l$ given above, we can estimate $r_Q\approx 10nm$ (see SI and Ref. \cite{laikhtman_exciton_2009} for the calculation of $r_Q$). Such a comparison yields $\langle r \rangle/2r_Q \approx 1.5$. This result suggests that the quantum mechanical motion of the dipoles due to the uncertainty principle (their zero-point motion) is of the order of the inter-particle distance. Two important properties of the liquid can be inferred from this ratio: the first is that quantum mechanical fluctuations prevent formation of long range order and crystallization. The second is that it might suggest that the IX liquid has strong quantum mechanical properties, similarly to liquid $^4He$ for example.  Indeed, the same calculation (see SI) shows that the kinetic energy due to the quantum motion $E_Q$ is much larger than the classical kinetic energy $k_BT$, and that it contributes about 25\% of the total liquid energy.

In conclusion, we observed phase transitions of a dipolar exciton fluid from a gas to a highly correlated dark liquid, and then to an e-h plasma, as the excitation power is increased at low enough temperatures. The liquid state, which results from a condensation of dark IX particles, has a \textit{well defined density}, indicating a significant reduction in its compressibility.
The analysis of the liquid's energy and other properties suggests that the inter-particle distance approximately corresponds to a close-packing of the dipolar excitons, and that the liquid might have a significant quantum uncertainty energy contribution, quite similar to liquid $^4He$. This is a new and exciting example of a collective state of a two-dimensional interacting system with internal spin degrees of freedom. Furthermore, this artificial atomic-like fluid resides inside a low-dimensional semiconductor structure, and it is a unique example where coupling of light to internal spin transitions plays a crucial role in the formation dynamics and in determining  the collective ground state properties.

\begin{acknowledgments}
We would like to acknowledge financial support from the German DFG (grant No. SA-598/9), from the German- Israeli Foundation (GIF I-1277-303.10/2014), and from the Israeli Science Foundation (grant No. 1319/12). The work at Princeton University was funded by the Gordon and Betty Moore Foundation through the EPiQS initiative Grant GBMF4420, and by the National Science Foundation MRSEC Grant DMR-1420541.
\end{acknowledgments}

\bibliographystyle{apsrev4-1}
\bibliography{exciton_lib_zotero}

\begin{thebibliography}{33}%
\makeatletter
\providecommand \@ifxundefined [1]{%
 \@ifx{#1\undefined}
}%
\providecommand \@ifnum [1]{%
 \ifnum #1\expandafter \@firstoftwo
 \else \expandafter \@secondoftwo
 \fi
}%
\providecommand \@ifx [1]{%
 \ifx #1\expandafter \@firstoftwo
 \else \expandafter \@secondoftwo
 \fi
}%
\providecommand \natexlab [1]{#1}%
\providecommand \enquote  [1]{``#1''}%
\providecommand \bibnamefont  [1]{#1}%
\providecommand \bibfnamefont [1]{#1}%
\providecommand \citenamefont [1]{#1}%
\providecommand \href@noop [0]{\@secondoftwo}%
\providecommand \href [0]{\begingroup \@sanitize@url \@href}%
\providecommand \@href[1]{\@@startlink{#1}\@@href}%
\providecommand \@@href[1]{\endgroup#1\@@endlink}%
\providecommand \@sanitize@url [0]{\catcode `\\12\catcode `\$12\catcode
  `\&12\catcode `\#12\catcode `\^12\catcode `\_12\catcode `\%12\relax}%
\providecommand \@@startlink[1]{}%
\providecommand \@@endlink[0]{}%
\providecommand \url  [0]{\begingroup\@sanitize@url \@url }%
\providecommand \@url [1]{\endgroup\@href {#1}{\urlprefix }}%
\providecommand \urlprefix  [0]{URL }%
\providecommand \Eprint [0]{\href }%
\providecommand \doibase [0]{http://dx.doi.org/}%
\providecommand \selectlanguage [0]{\@gobble}%
\providecommand \bibinfo  [0]{\@secondoftwo}%
\providecommand \bibfield  [0]{\@secondoftwo}%
\providecommand \translation [1]{[#1]}%
\providecommand \BibitemOpen [0]{}%
\providecommand \bibitemStop [0]{}%
\providecommand \bibitemNoStop [0]{.\EOS\space}%
\providecommand \EOS [0]{\spacefactor3000\relax}%
\providecommand \BibitemShut  [1]{\csname bibitem#1\endcsname}%
\let\auto@bib@innerbib\@empty
\bibitem [{\citenamefont {Santos}\ \emph {et~al.}(2000)\citenamefont {Santos},
  \citenamefont {Shlyapnikov}, \citenamefont {Zoller},\ and\ \citenamefont
  {Lewenstein}}]{santos_bose-einstein_2000}%
  \BibitemOpen
  \bibfield  {author} {\bibinfo {author} {\bibfnamefont {L.}~\bibnamefont
  {Santos}}, \bibinfo {author} {\bibfnamefont {G.~V.}\ \bibnamefont
  {Shlyapnikov}}, \bibinfo {author} {\bibfnamefont {P.}~\bibnamefont {Zoller}},
  \ and\ \bibinfo {author} {\bibfnamefont {M.}~\bibnamefont {Lewenstein}},\
  }\href {\doibase 10.1103/PhysRevLett.85.1791} {\bibfield  {journal} {\bibinfo
   {journal} {Phys. Rev. Lett.}\ }\textbf {\bibinfo {volume} {85}},\ \bibinfo
  {pages} {1791} (\bibinfo {year} {2000})}\BibitemShut {NoStop}%
\bibitem [{\citenamefont {B\"uchler}\ \emph {et~al.}(2007)\citenamefont
  {B\"uchler}, \citenamefont {Demler}, \citenamefont {Lukin}, \citenamefont
  {Micheli}, \citenamefont {Prokof'ev}, \citenamefont {Pupillo},\ and\
  \citenamefont {Zoller}}]{buchler_strongly_2007}%
  \BibitemOpen
  \bibfield  {author} {\bibinfo {author} {\bibfnamefont {H.~P.}\ \bibnamefont
  {B\"uchler}}, \bibinfo {author} {\bibfnamefont {E.}~\bibnamefont {Demler}},
  \bibinfo {author} {\bibfnamefont {M.}~\bibnamefont {Lukin}}, \bibinfo
  {author} {\bibfnamefont {A.}~\bibnamefont {Micheli}}, \bibinfo {author}
  {\bibfnamefont {N.}~\bibnamefont {Prokof'ev}}, \bibinfo {author}
  {\bibfnamefont {G.}~\bibnamefont {Pupillo}}, \ and\ \bibinfo {author}
  {\bibfnamefont {P.}~\bibnamefont {Zoller}},\ }\href {\doibase
  10.1103/PhysRevLett.98.060404} {\bibfield  {journal} {\bibinfo  {journal}
  {Phys. Rev. Lett.}\ }\textbf {\bibinfo {volume} {98}},\ \bibinfo {pages}
  {060404} (\bibinfo {year} {2007})}\BibitemShut {NoStop}%
\bibitem [{\citenamefont {Lahaye}\ \emph {et~al.}(2009)\citenamefont {Lahaye},
  \citenamefont {Menotti}, \citenamefont {Santos}, \citenamefont {Lewenstein},\
  and\ \citenamefont {Pfau}}]{lahaye_physics_2009}%
  \BibitemOpen
  \bibfield  {author} {\bibinfo {author} {\bibfnamefont {T.}~\bibnamefont
  {Lahaye}}, \bibinfo {author} {\bibfnamefont {C.}~\bibnamefont {Menotti}},
  \bibinfo {author} {\bibfnamefont {L.}~\bibnamefont {Santos}}, \bibinfo
  {author} {\bibfnamefont {M.}~\bibnamefont {Lewenstein}}, \ and\ \bibinfo
  {author} {\bibfnamefont {T.}~\bibnamefont {Pfau}},\ }\href {\doibase
  10.1088/0034-4885/72/12/126401} {\bibfield  {journal} {\bibinfo  {journal}
  {Rep. Prog. Phys.}\ }\textbf {\bibinfo {volume} {72}},\ \bibinfo {pages}
  {126401} (\bibinfo {year} {2009})}\BibitemShut {NoStop}%
\bibitem [{\citenamefont {Astrakharchik}\ \emph {et~al.}(2007)\citenamefont
  {Astrakharchik}, \citenamefont {Boronat}, \citenamefont {Kurbakov},\ and\
  \citenamefont {Lozovik}}]{astrakharchik_quantum_2007}%
  \BibitemOpen
  \bibfield  {author} {\bibinfo {author} {\bibfnamefont {G.~E.}\ \bibnamefont
  {Astrakharchik}}, \bibinfo {author} {\bibfnamefont {J.}~\bibnamefont
  {Boronat}}, \bibinfo {author} {\bibfnamefont {I.~L.}\ \bibnamefont
  {Kurbakov}}, \ and\ \bibinfo {author} {\bibfnamefont {Y.~E.}\ \bibnamefont
  {Lozovik}},\ }\href {\doibase 10.1103/PhysRevLett.98.060405} {\bibfield
  {journal} {\bibinfo  {journal} {Phys. Rev. Lett.}\ }\textbf {\bibinfo
  {volume} {98}},\ \bibinfo {pages} {060405} (\bibinfo {year}
  {2007})}\BibitemShut {NoStop}%
\bibitem [{\citenamefont {Laikhtman}\ and\ \citenamefont
  {Rapaport}(2009{\natexlab{a}})}]{laikhtman_correlations_2009}%
  \BibitemOpen
  \bibfield  {author} {\bibinfo {author} {\bibfnamefont {B.}~\bibnamefont
  {Laikhtman}}\ and\ \bibinfo {author} {\bibfnamefont {R.}~\bibnamefont
  {Rapaport}},\ }\href {\doibase 10.1209/0295-5075/87/27010} {\bibfield
  {journal} {\bibinfo  {journal} {{EPL}}\ }\textbf {\bibinfo {volume} {87}},\
  \bibinfo {pages} {27010} (\bibinfo {year} {2009}{\natexlab{a}})}\BibitemShut
  {NoStop}%
\bibitem [{\citenamefont {Laikhtman}\ and\ \citenamefont
  {Rapaport}(2009{\natexlab{b}})}]{laikhtman_exciton_2009}%
  \BibitemOpen
  \bibfield  {author} {\bibinfo {author} {\bibfnamefont {B.}~\bibnamefont
  {Laikhtman}}\ and\ \bibinfo {author} {\bibfnamefont {R.}~\bibnamefont
  {Rapaport}},\ }\href {\doibase 10.1103/PhysRevB.80.195313} {\bibfield
  {journal} {\bibinfo  {journal} {Phys. Rev. B}\ }\textbf {\bibinfo {volume}
  {80}},\ \bibinfo {pages} {195313} (\bibinfo {year}
  {2009}{\natexlab{b}})}\BibitemShut {NoStop}%
\bibitem [{\citenamefont {Moroni}\ and\ \citenamefont
  {Boninsegni}(2014)}]{moroni_coexistence_2014}%
  \BibitemOpen
  \bibfield  {author} {\bibinfo {author} {\bibfnamefont {S.}~\bibnamefont
  {Moroni}}\ and\ \bibinfo {author} {\bibfnamefont {M.}~\bibnamefont
  {Boninsegni}},\ }\href {\doibase 10.1103/PhysRevLett.113.240407} {\bibfield
  {journal} {\bibinfo  {journal} {Phys. Rev. Lett.}\ }\textbf {\bibinfo
  {volume} {113}},\ \bibinfo {pages} {240407} (\bibinfo {year}
  {2014})}\BibitemShut {NoStop}%
\bibitem [{\citenamefont {Lu}\ \emph {et~al.}(2011)\citenamefont {Lu},
  \citenamefont {Burdick}, \citenamefont {Youn},\ and\ \citenamefont
  {Lev}}]{lu_strongly_2011}%
  \BibitemOpen
  \bibfield  {author} {\bibinfo {author} {\bibfnamefont {M.}~\bibnamefont
  {Lu}}, \bibinfo {author} {\bibfnamefont {N.~Q.}\ \bibnamefont {Burdick}},
  \bibinfo {author} {\bibfnamefont {S.~H.}\ \bibnamefont {Youn}}, \ and\
  \bibinfo {author} {\bibfnamefont {B.~L.}\ \bibnamefont {Lev}},\ }\href
  {\doibase 10.1103/PhysRevLett.107.190401} {\bibfield  {journal} {\bibinfo
  {journal} {Phys. Rev. Lett.}\ }\textbf {\bibinfo {volume} {107}},\ \bibinfo
  {pages} {190401} (\bibinfo {year} {2011})}\BibitemShut {NoStop}%
\bibitem [{\citenamefont {Gaj}\ \emph {et~al.}(2014)\citenamefont {Gaj},
  \citenamefont {Krupp}, \citenamefont {Balewski}, \citenamefont {L{\"o}w},
  \citenamefont {Hofferberth},\ and\ \citenamefont
  {Pfau}}]{gaj_molecular_2014}%
  \BibitemOpen
  \bibfield  {author} {\bibinfo {author} {\bibfnamefont {A.}~\bibnamefont
  {Gaj}}, \bibinfo {author} {\bibfnamefont {A.~T.}\ \bibnamefont {Krupp}},
  \bibinfo {author} {\bibfnamefont {J.~B.}\ \bibnamefont {Balewski}}, \bibinfo
  {author} {\bibfnamefont {R.}~\bibnamefont {L{\"o}w}}, \bibinfo {author}
  {\bibfnamefont {S.}~\bibnamefont {Hofferberth}}, \ and\ \bibinfo {author}
  {\bibfnamefont {T.}~\bibnamefont {Pfau}},\ }\href {\doibase
  10.1038/ncomms5546} {\bibfield  {journal} {\bibinfo  {journal} {Nat Commun}\
  }\textbf {\bibinfo {volume} {5}} (\bibinfo {year} {2014}),\
  10.1038/ncomms5546}\BibitemShut {NoStop}%
\bibitem [{\citenamefont {Firstenberg}\ \emph {et~al.}(2013)\citenamefont
  {Firstenberg}, \citenamefont {Peyronel}, \citenamefont {Liang}, \citenamefont
  {Gorshkov}, \citenamefont {Lukin},\ and\ \citenamefont
  {Vuleti{\'c}}}]{firstenberg_attractive_2013}%
  \BibitemOpen
  \bibfield  {author} {\bibinfo {author} {\bibfnamefont {O.}~\bibnamefont
  {Firstenberg}}, \bibinfo {author} {\bibfnamefont {T.}~\bibnamefont
  {Peyronel}}, \bibinfo {author} {\bibfnamefont {Q.-Y.}\ \bibnamefont {Liang}},
  \bibinfo {author} {\bibfnamefont {A.~V.}\ \bibnamefont {Gorshkov}}, \bibinfo
  {author} {\bibfnamefont {M.~D.}\ \bibnamefont {Lukin}}, \ and\ \bibinfo
  {author} {\bibfnamefont {V.}~\bibnamefont {Vuleti{\'c}}},\ }\href {\doibase
  10.1038/nature12512} {\bibfield  {journal} {\bibinfo  {journal} {Nature}\
  }\textbf {\bibinfo {volume} {502}},\ \bibinfo {pages} {71} (\bibinfo {year}
  {2013})}\BibitemShut {NoStop}%
\bibitem [{\citenamefont {Jin}\ and\ \citenamefont
  {Ye}(2011)}]{jin_polar_2011}%
  \BibitemOpen
  \bibfield  {author} {\bibinfo {author} {\bibfnamefont {D.~S.}\ \bibnamefont
  {Jin}}\ and\ \bibinfo {author} {\bibfnamefont {J.}~\bibnamefont {Ye}},\
  }\href {\doibase 10.1063/1.3592002} {\bibfield  {journal} {\bibinfo
  {journal} {Physics Today}\ }\textbf {\bibinfo {volume} {64}},\ \bibinfo
  {pages} {27} (\bibinfo {year} {2011})}\BibitemShut {NoStop}%
\bibitem [{\citenamefont {Yan}\ \emph {et~al.}(2013)\citenamefont {Yan},
  \citenamefont {Moses}, \citenamefont {Gadway}, \citenamefont {Covey},
  \citenamefont {Hazzard}, \citenamefont {Rey}, \citenamefont {Jin},\ and\
  \citenamefont {Ye}}]{yan_observation_2013}%
  \BibitemOpen
  \bibfield  {author} {\bibinfo {author} {\bibfnamefont {B.}~\bibnamefont
  {Yan}}, \bibinfo {author} {\bibfnamefont {S.~A.}\ \bibnamefont {Moses}},
  \bibinfo {author} {\bibfnamefont {B.}~\bibnamefont {Gadway}}, \bibinfo
  {author} {\bibfnamefont {J.~P.}\ \bibnamefont {Covey}}, \bibinfo {author}
  {\bibfnamefont {K.~R.~A.}\ \bibnamefont {Hazzard}}, \bibinfo {author}
  {\bibfnamefont {A.~M.}\ \bibnamefont {Rey}}, \bibinfo {author} {\bibfnamefont
  {D.~S.}\ \bibnamefont {Jin}}, \ and\ \bibinfo {author} {\bibfnamefont
  {J.}~\bibnamefont {Ye}},\ }\href {\doibase 10.1038/nature12483} {\bibfield
  {journal} {\bibinfo  {journal} {Nature}\ }\textbf {\bibinfo {volume} {501}},\
  \bibinfo {pages} {521} (\bibinfo {year} {2013})}\BibitemShut {NoStop}%
\bibitem [{\citenamefont {Eisenstein}\ and\ \citenamefont
  {MacDonald}(2004)}]{eisenstein_bose-einstein_2004}%
  \BibitemOpen
  \bibfield  {author} {\bibinfo {author} {\bibfnamefont {J.~P.}\ \bibnamefont
  {Eisenstein}}\ and\ \bibinfo {author} {\bibfnamefont {A.~H.}\ \bibnamefont
  {MacDonald}},\ }\href {\doibase 10.1038/nature03081} {\bibfield  {journal}
  {\bibinfo  {journal} {Nature}\ }\textbf {\bibinfo {volume} {432}},\ \bibinfo
  {pages} {691} (\bibinfo {year} {2004})}\BibitemShut {NoStop}%
\bibitem [{\citenamefont {High}\ \emph
  {et~al.}(2012{\natexlab{a}})\citenamefont {High}, \citenamefont {Leonard},
  \citenamefont {Hammack}, \citenamefont {Fogler}, \citenamefont {Butov},
  \citenamefont {Kavokin}, \citenamefont {Campman},\ and\ \citenamefont
  {Gossard}}]{high_spontaneous_2012}%
  \BibitemOpen
  \bibfield  {author} {\bibinfo {author} {\bibfnamefont {A.~A.}\ \bibnamefont
  {High}}, \bibinfo {author} {\bibfnamefont {J.~R.}\ \bibnamefont {Leonard}},
  \bibinfo {author} {\bibfnamefont {A.~T.}\ \bibnamefont {Hammack}}, \bibinfo
  {author} {\bibfnamefont {M.~M.}\ \bibnamefont {Fogler}}, \bibinfo {author}
  {\bibfnamefont {L.~V.}\ \bibnamefont {Butov}}, \bibinfo {author}
  {\bibfnamefont {A.~V.}\ \bibnamefont {Kavokin}}, \bibinfo {author}
  {\bibfnamefont {K.~L.}\ \bibnamefont {Campman}}, \ and\ \bibinfo {author}
  {\bibfnamefont {A.~C.}\ \bibnamefont {Gossard}},\ }\href {\doibase
  10.1038/nature10903} {\bibfield  {journal} {\bibinfo  {journal} {Nature}\
  }\textbf {\bibinfo {volume} {483}},\ \bibinfo {pages} {584} (\bibinfo {year}
  {2012}{\natexlab{a}})}\BibitemShut {NoStop}%
\bibitem [{\citenamefont {Shilo}\ \emph {et~al.}(2013)\citenamefont {Shilo},
  \citenamefont {Cohen}, \citenamefont {Laikhtman}, \citenamefont {West},
  \citenamefont {Pfeiffer},\ and\ \citenamefont
  {Rapaport}}]{shilo_particle_2013}%
  \BibitemOpen
  \bibfield  {author} {\bibinfo {author} {\bibfnamefont {Y.}~\bibnamefont
  {Shilo}}, \bibinfo {author} {\bibfnamefont {K.}~\bibnamefont {Cohen}},
  \bibinfo {author} {\bibfnamefont {B.}~\bibnamefont {Laikhtman}}, \bibinfo
  {author} {\bibfnamefont {K.}~\bibnamefont {West}}, \bibinfo {author}
  {\bibfnamefont {L.}~\bibnamefont {Pfeiffer}}, \ and\ \bibinfo {author}
  {\bibfnamefont {R.}~\bibnamefont {Rapaport}},\ }\href {\doibase
  10.1038/ncomms3335} {\bibfield  {journal} {\bibinfo  {journal} {Nat Commun}\
  }\textbf {\bibinfo {volume} {4}} (\bibinfo {year} {2013}),\
  10.1038/ncomms3335}\BibitemShut {NoStop}%
\bibitem [{\citenamefont {Kazimierczuk}\ \emph {et~al.}(2014)\citenamefont
  {Kazimierczuk}, \citenamefont {Fr{\"o}hlich}, \citenamefont {Scheel},
  \citenamefont {Stolz},\ and\ \citenamefont
  {Bayer}}]{kazimierczuk_giant_2014}%
  \BibitemOpen
  \bibfield  {author} {\bibinfo {author} {\bibfnamefont {T.}~\bibnamefont
  {Kazimierczuk}}, \bibinfo {author} {\bibfnamefont {D.}~\bibnamefont
  {Fr{\"o}hlich}}, \bibinfo {author} {\bibfnamefont {S.}~\bibnamefont
  {Scheel}}, \bibinfo {author} {\bibfnamefont {H.}~\bibnamefont {Stolz}}, \
  and\ \bibinfo {author} {\bibfnamefont {M.}~\bibnamefont {Bayer}},\ }\href
  {\doibase 10.1038/nature13832} {\bibfield  {journal} {\bibinfo  {journal}
  {Nature}\ }\textbf {\bibinfo {volume} {514}},\ \bibinfo {pages} {343}
  (\bibinfo {year} {2014})}\BibitemShut {NoStop}%
\bibitem [{\citenamefont {Butov}(2004)}]{butov_condensation_2004}%
  \BibitemOpen
  \bibfield  {author} {\bibinfo {author} {\bibfnamefont {L.~V.}\ \bibnamefont
  {Butov}},\ }\href {\doibase 10.1088/0953-8984/16/50/R02} {\bibfield
  {journal} {\bibinfo  {journal} {J. Phys.: Condens. Matter}\ }\textbf
  {\bibinfo {volume} {16}},\ \bibinfo {pages} {R1577} (\bibinfo {year}
  {2004})}\BibitemShut {NoStop}%
\bibitem [{\citenamefont {Rapaport}\ and\ \citenamefont
  {Chen}(2007)}]{rapaport_experimental_2007}%
  \BibitemOpen
  \bibfield  {author} {\bibinfo {author} {\bibfnamefont {R.}~\bibnamefont
  {Rapaport}}\ and\ \bibinfo {author} {\bibfnamefont {G.}~\bibnamefont
  {Chen}},\ }\href {\doibase 10.1088/0953-8984/19/29/295207} {\bibfield
  {journal} {\bibinfo  {journal} {J. Phys.: Condens. Matter}\ }\textbf
  {\bibinfo {volume} {19}},\ \bibinfo {pages} {295207} (\bibinfo {year}
  {2007})}\BibitemShut {NoStop}%
\bibitem [{\citenamefont {Poem}\ \emph {et~al.}(2010)\citenamefont {Poem},
  \citenamefont {Kodriano}, \citenamefont {Tradonsky}, \citenamefont {Lindner},
  \citenamefont {Gerardot}, \citenamefont {Petroff},\ and\ \citenamefont
  {Gershoni}}]{poem_accessing_2010}%
  \BibitemOpen
  \bibfield  {author} {\bibinfo {author} {\bibfnamefont {E.}~\bibnamefont
  {Poem}}, \bibinfo {author} {\bibfnamefont {Y.}~\bibnamefont {Kodriano}},
  \bibinfo {author} {\bibfnamefont {C.}~\bibnamefont {Tradonsky}}, \bibinfo
  {author} {\bibfnamefont {N.~H.}\ \bibnamefont {Lindner}}, \bibinfo {author}
  {\bibfnamefont {B.~D.}\ \bibnamefont {Gerardot}}, \bibinfo {author}
  {\bibfnamefont {P.~M.}\ \bibnamefont {Petroff}}, \ and\ \bibinfo {author}
  {\bibfnamefont {D.}~\bibnamefont {Gershoni}},\ }\href {\doibase
  10.1038/nphys1812} {\bibfield  {journal} {\bibinfo  {journal} {Nat Phys}\
  }\textbf {\bibinfo {volume} {6}},\ \bibinfo {pages} {993} (\bibinfo {year}
  {2010})}\BibitemShut {NoStop}%
\bibitem [{\citenamefont {Maialle}\ \emph {et~al.}(1993)\citenamefont
  {Maialle}, \citenamefont {de~Andrada~e Silva},\ and\ \citenamefont
  {Sham}}]{maialle_exciton_1993}%
  \BibitemOpen
  \bibfield  {author} {\bibinfo {author} {\bibfnamefont {M.~Z.}\ \bibnamefont
  {Maialle}}, \bibinfo {author} {\bibfnamefont {E.~A.}\ \bibnamefont
  {de~Andrada~e Silva}}, \ and\ \bibinfo {author} {\bibfnamefont {L.~J.}\
  \bibnamefont {Sham}},\ }\href {\doibase 10.1103/PhysRevB.47.15776} {\bibfield
   {journal} {\bibinfo  {journal} {Phys. Rev. B}\ }\textbf {\bibinfo {volume}
  {47}},\ \bibinfo {pages} {15776} (\bibinfo {year} {1993})}\BibitemShut
  {NoStop}%
\bibitem [{\citenamefont {Combescot}\ \emph {et~al.}(2007)\citenamefont
  {Combescot}, \citenamefont {Betbeder-Matibet},\ and\ \citenamefont
  {Combescot}}]{combescot_bose-einstein_2007}%
  \BibitemOpen
  \bibfield  {author} {\bibinfo {author} {\bibfnamefont {M.}~\bibnamefont
  {Combescot}}, \bibinfo {author} {\bibfnamefont {O.}~\bibnamefont
  {Betbeder-Matibet}}, \ and\ \bibinfo {author} {\bibfnamefont
  {R.}~\bibnamefont {Combescot}},\ }\href {\doibase
  10.1103/PhysRevLett.99.176403} {\bibfield  {journal} {\bibinfo  {journal}
  {Phys. Rev. Lett.}\ }\textbf {\bibinfo {volume} {99}},\ \bibinfo {pages}
  {176403} (\bibinfo {year} {2007})}\BibitemShut {NoStop}%
\bibitem [{\citenamefont {High}\ \emph
  {et~al.}(2012{\natexlab{b}})\citenamefont {High}, \citenamefont {Leonard},
  \citenamefont {Remeika}, \citenamefont {Butov}, \citenamefont {Hanson},\ and\
  \citenamefont {Gossard}}]{high_condensation_2012}%
  \BibitemOpen
  \bibfield  {author} {\bibinfo {author} {\bibfnamefont {A.~A.}\ \bibnamefont
  {High}}, \bibinfo {author} {\bibfnamefont {J.~R.}\ \bibnamefont {Leonard}},
  \bibinfo {author} {\bibfnamefont {M.}~\bibnamefont {Remeika}}, \bibinfo
  {author} {\bibfnamefont {L.~V.}\ \bibnamefont {Butov}}, \bibinfo {author}
  {\bibfnamefont {M.}~\bibnamefont {Hanson}}, \ and\ \bibinfo {author}
  {\bibfnamefont {A.~C.}\ \bibnamefont {Gossard}},\ }\href {\doibase
  10.1021/nl300983n} {\bibfield  {journal} {\bibinfo  {journal} {Nano Lett.}\
  }\textbf {\bibinfo {volume} {12}},\ \bibinfo {pages} {2605} (\bibinfo {year}
  {2012}{\natexlab{b}})}\BibitemShut {NoStop}%
\bibitem [{\citenamefont {High}\ \emph {et~al.}(2013)\citenamefont {High},
  \citenamefont {Hammack}, \citenamefont {Leonard}, \citenamefont {Yang},
  \citenamefont {Butov}, \citenamefont {Ostatnick\'{y}}, \citenamefont
  {Vladimirova}, \citenamefont {Kavokin}, \citenamefont {Liew}, \citenamefont
  {Campman},\ and\ \citenamefont {Gossard}}]{high_spin_2013}%
  \BibitemOpen
  \bibfield  {author} {\bibinfo {author} {\bibfnamefont {A.~A.}\ \bibnamefont
  {High}}, \bibinfo {author} {\bibfnamefont {A.~T.}\ \bibnamefont {Hammack}},
  \bibinfo {author} {\bibfnamefont {J.~R.}\ \bibnamefont {Leonard}}, \bibinfo
  {author} {\bibfnamefont {S.}~\bibnamefont {Yang}}, \bibinfo {author}
  {\bibfnamefont {L.~V.}\ \bibnamefont {Butov}}, \bibinfo {author}
  {\bibfnamefont {T.}~\bibnamefont {Ostatnick\'{y}}}, \bibinfo {author}
  {\bibfnamefont {M.}~\bibnamefont {Vladimirova}}, \bibinfo {author}
  {\bibfnamefont {A.~V.}\ \bibnamefont {Kavokin}}, \bibinfo {author}
  {\bibfnamefont {T.~C.~H.}\ \bibnamefont {Liew}}, \bibinfo {author}
  {\bibfnamefont {K.~L.}\ \bibnamefont {Campman}}, \ and\ \bibinfo {author}
  {\bibfnamefont {A.~C.}\ \bibnamefont {Gossard}},\ }\href {\doibase
  10.1103/PhysRevLett.110.246403} {\bibfield  {journal} {\bibinfo  {journal}
  {Phys. Rev. Lett.}\ }\textbf {\bibinfo {volume} {110}},\ \bibinfo {pages}
  {246403} (\bibinfo {year} {2013})}\BibitemShut {NoStop}%
\bibitem [{\citenamefont {Alloing}\ \emph {et~al.}(2014)\citenamefont
  {Alloing}, \citenamefont {Beian}, \citenamefont {Lewenstein}, \citenamefont
  {Fuster}, \citenamefont {Gonz{\'a}lez}, \citenamefont {Gonz{\'a}lez},
  \citenamefont {Combescot}, \citenamefont {Combescot},\ and\ \citenamefont
  {Dubin}}]{alloing_evidence_2014}%
  \BibitemOpen
  \bibfield  {author} {\bibinfo {author} {\bibfnamefont {M.}~\bibnamefont
  {Alloing}}, \bibinfo {author} {\bibfnamefont {M.}~\bibnamefont {Beian}},
  \bibinfo {author} {\bibfnamefont {M.}~\bibnamefont {Lewenstein}}, \bibinfo
  {author} {\bibfnamefont {D.}~\bibnamefont {Fuster}}, \bibinfo {author}
  {\bibfnamefont {Y.}~\bibnamefont {Gonz{\'a}lez}}, \bibinfo {author}
  {\bibfnamefont {L.}~\bibnamefont {Gonz{\'a}lez}}, \bibinfo {author}
  {\bibfnamefont {R.}~\bibnamefont {Combescot}}, \bibinfo {author}
  {\bibfnamefont {M.}~\bibnamefont {Combescot}}, \ and\ \bibinfo {author}
  {\bibfnamefont {F.}~\bibnamefont {Dubin}},\ }\href {\doibase
  10.1209/0295-5075/107/10012} {\bibfield  {journal} {\bibinfo  {journal}
  {{EPL}}\ }\textbf {\bibinfo {volume} {107}},\ \bibinfo {pages} {10012}
  (\bibinfo {year} {2014})}\BibitemShut {NoStop}%
\bibitem [{\citenamefont {Stern}\ \emph {et~al.}(2014)\citenamefont {Stern},
  \citenamefont {Umansky},\ and\ \citenamefont
  {Bar-Joseph}}]{stern_exciton_2014}%
  \BibitemOpen
  \bibfield  {author} {\bibinfo {author} {\bibfnamefont {M.}~\bibnamefont
  {Stern}}, \bibinfo {author} {\bibfnamefont {V.}~\bibnamefont {Umansky}}, \
  and\ \bibinfo {author} {\bibfnamefont {I.}~\bibnamefont {Bar-Joseph}},\
  }\href {\doibase 10.1126/science.1243409} {\bibfield  {journal} {\bibinfo
  {journal} {Science}\ }\textbf {\bibinfo {volume} {343}},\ \bibinfo {pages}
  {55} (\bibinfo {year} {2014})}\BibitemShut {NoStop}%
\bibitem [{\citenamefont {Rapaport}\ \emph {et~al.}(2005)\citenamefont
  {Rapaport}, \citenamefont {Chen}, \citenamefont {Simon}, \citenamefont
  {Mitrofanov}, \citenamefont {Pfeiffer},\ and\ \citenamefont
  {Platzman}}]{rapaport_electrostatic_2005}%
  \BibitemOpen
  \bibfield  {author} {\bibinfo {author} {\bibfnamefont {R.}~\bibnamefont
  {Rapaport}}, \bibinfo {author} {\bibfnamefont {G.}~\bibnamefont {Chen}},
  \bibinfo {author} {\bibfnamefont {S.}~\bibnamefont {Simon}}, \bibinfo
  {author} {\bibfnamefont {O.}~\bibnamefont {Mitrofanov}}, \bibinfo {author}
  {\bibfnamefont {L.}~\bibnamefont {Pfeiffer}}, \ and\ \bibinfo {author}
  {\bibfnamefont {P.~M.}\ \bibnamefont {Platzman}},\ }\href {\doibase
  10.1103/PhysRevB.72.075428} {\bibfield  {journal} {\bibinfo  {journal} {Phys.
  Rev. B}\ }\textbf {\bibinfo {volume} {72}},\ \bibinfo {pages} {075428}
  (\bibinfo {year} {2005})}\BibitemShut {NoStop}%
\bibitem [{\citenamefont {Shilo}(2014)}]{shilo_optical_2014}%
  \BibitemOpen
  \bibfield  {author} {\bibinfo {author} {\bibfnamefont {Y.}~\bibnamefont
  {Shilo}},\ }\emph {\bibinfo {title} {Optical study of two dimensional dipolar
  exciton fluids}},\ \href {http://hufind.huji.ac.il/Record/HUJ001973786}
  {Ph.D. thesis},\ \bibinfo  {school} {Hebrew University of Jerusalem}
  (\bibinfo {year} {2014})\BibitemShut {NoStop}%
\bibitem [{\citenamefont {Zimmermann}\ and\ \citenamefont
  {Schindler}(2007)}]{zimmermann_excitonexciton_2007}%
  \BibitemOpen
  \bibfield  {author} {\bibinfo {author} {\bibfnamefont {R.}~\bibnamefont
  {Zimmermann}}\ and\ \bibinfo {author} {\bibfnamefont {C.}~\bibnamefont
  {Schindler}},\ }\href {\doibase 10.1016/j.ssc.2007.07.044} {\bibfield
  {journal} {\bibinfo  {journal} {Solid State Communications}\ }\textbf
  {\bibinfo {volume} {144}},\ \bibinfo {pages} {395} (\bibinfo {year}
  {2007})}\BibitemShut {NoStop}%
\bibitem [{Note1()}]{Note1}%
  \BibitemOpen
  \bibinfo {note} {This increase cannot be due to a sudden increase in $C$,
  since particles at lower temperatures have lower kinetic energies, higher
  spatial correlation, and thus smaller values of $C$}\BibitemShut {NoStop}%
\bibitem [{\citenamefont {Lee}\ \emph {et~al.}(2009)\citenamefont {Lee},
  \citenamefont {Drummond},\ and\ \citenamefont
  {Needs}}]{lee_exciton-exciton_2009}%
  \BibitemOpen
  \bibfield  {author} {\bibinfo {author} {\bibfnamefont {R.~M.}\ \bibnamefont
  {Lee}}, \bibinfo {author} {\bibfnamefont {N.~D.}\ \bibnamefont {Drummond}}, \
  and\ \bibinfo {author} {\bibfnamefont {R.~J.}\ \bibnamefont {Needs}},\ }\href
  {\doibase 10.1103/PhysRevB.79.125308} {\bibfield  {journal} {\bibinfo
  {journal} {Phys. Rev. B}\ }\textbf {\bibinfo {volume} {79}},\ \bibinfo
  {pages} {125308} (\bibinfo {year} {2009})}\BibitemShut {NoStop}%
\bibitem [{Note2()}]{Note2}%
  \BibitemOpen
  \bibinfo {note} {We note however that this estimate is for a non-interacting
  Bose gas, while as mentioned in Ref.\cite {laikhtman_exciton_2009} quantum
  degeneracy could be suppressed down to lower temperatures due to
  multi-particle correlations}\BibitemShut {NoStop}%
\bibitem [{\citenamefont {Kasprzak}\ \emph {et~al.}(2006)\citenamefont
  {Kasprzak}, \citenamefont {Richard}, \citenamefont {Kundermann},
  \citenamefont {Baas}, \citenamefont {Jeambrun}, \citenamefont {Keeling},
  \citenamefont {Marchetti}, \citenamefont {Szymanska}, \citenamefont {Andre},
  \citenamefont {Staehli}, \citenamefont {Savona}, \citenamefont {Littlewood},
  \citenamefont {Deveaud},\ and\ \citenamefont
  {Dang}}]{kasprzak_bose-einstein_2006}%
  \BibitemOpen
  \bibfield  {author} {\bibinfo {author} {\bibfnamefont {J.}~\bibnamefont
  {Kasprzak}}, \bibinfo {author} {\bibfnamefont {M.}~\bibnamefont {Richard}},
  \bibinfo {author} {\bibfnamefont {S.}~\bibnamefont {Kundermann}}, \bibinfo
  {author} {\bibfnamefont {A.}~\bibnamefont {Baas}}, \bibinfo {author}
  {\bibfnamefont {P.}~\bibnamefont {Jeambrun}}, \bibinfo {author}
  {\bibfnamefont {J.~M.~J.}\ \bibnamefont {Keeling}}, \bibinfo {author}
  {\bibfnamefont {F.~M.}\ \bibnamefont {Marchetti}}, \bibinfo {author}
  {\bibfnamefont {M.~H.}\ \bibnamefont {Szymanska}}, \bibinfo {author}
  {\bibfnamefont {R.}~\bibnamefont {Andre}}, \bibinfo {author} {\bibfnamefont
  {J.~L.}\ \bibnamefont {Staehli}}, \bibinfo {author} {\bibfnamefont
  {V.}~\bibnamefont {Savona}}, \bibinfo {author} {\bibfnamefont {P.~B.}\
  \bibnamefont {Littlewood}}, \bibinfo {author} {\bibfnamefont
  {B.}~\bibnamefont {Deveaud}}, \ and\ \bibinfo {author} {\bibfnamefont
  {L.~S.}\ \bibnamefont {Dang}},\ }\href {\doibase 10.1038/nature05131}
  {\bibfield  {journal} {\bibinfo  {journal} {Nature}\ }\textbf {\bibinfo
  {volume} {443}},\ \bibinfo {pages} {409} (\bibinfo {year}
  {2006})}\BibitemShut {NoStop}%
\bibitem [{\citenamefont {Repp}\ \emph {et~al.}(2014)\citenamefont {Repp},
  \citenamefont {Schinner}, \citenamefont {Schubert}, \citenamefont {Rai},
  \citenamefont {Reuter}, \citenamefont {Wieck}, \citenamefont {Wurstbauer},
  \citenamefont {Kotthaus},\ and\ \citenamefont
  {Holleitner}}]{repp_confocal_2014}%
  \BibitemOpen
  \bibfield  {author} {\bibinfo {author} {\bibfnamefont {J.}~\bibnamefont
  {Repp}}, \bibinfo {author} {\bibfnamefont {G.~J.}\ \bibnamefont {Schinner}},
  \bibinfo {author} {\bibfnamefont {E.}~\bibnamefont {Schubert}}, \bibinfo
  {author} {\bibfnamefont {A.~K.}\ \bibnamefont {Rai}}, \bibinfo {author}
  {\bibfnamefont {D.}~\bibnamefont {Reuter}}, \bibinfo {author} {\bibfnamefont
  {A.~D.}\ \bibnamefont {Wieck}}, \bibinfo {author} {\bibfnamefont
  {U.}~\bibnamefont {Wurstbauer}}, \bibinfo {author} {\bibfnamefont {J.~P.}\
  \bibnamefont {Kotthaus}}, \ and\ \bibinfo {author} {\bibfnamefont {A.~W.}\
  \bibnamefont {Holleitner}},\ }\href {\doibase 10.1063/1.4904222} {\bibfield
  {journal} {\bibinfo  {journal} {Applied Physics Letters}\ }\textbf {\bibinfo
  {volume} {105}},\ \bibinfo {pages} {241101} (\bibinfo {year}
  {2014})}\BibitemShut {NoStop}%
\end{thebibliography}%




\pagebreak
\widetext
\graphicspath{{./SI_figs/}}

\begin{center}
\textbf{\large Supplementary Information}
\end{center}

\setcounter{equation}{0}
\setcounter{figure}{0}
\setcounter{table}{0}
\makeatletter
\renewcommand{\theequation}{S\arabic{equation}}
\renewcommand{\thefigure}{S\arabic{figure}}

\section*{Supplementary Note 1 - Raw data for different temperatures and excitation powers}
\begin{figure}[ht!]
  \centering
  \includegraphics[width=0.75\textwidth]{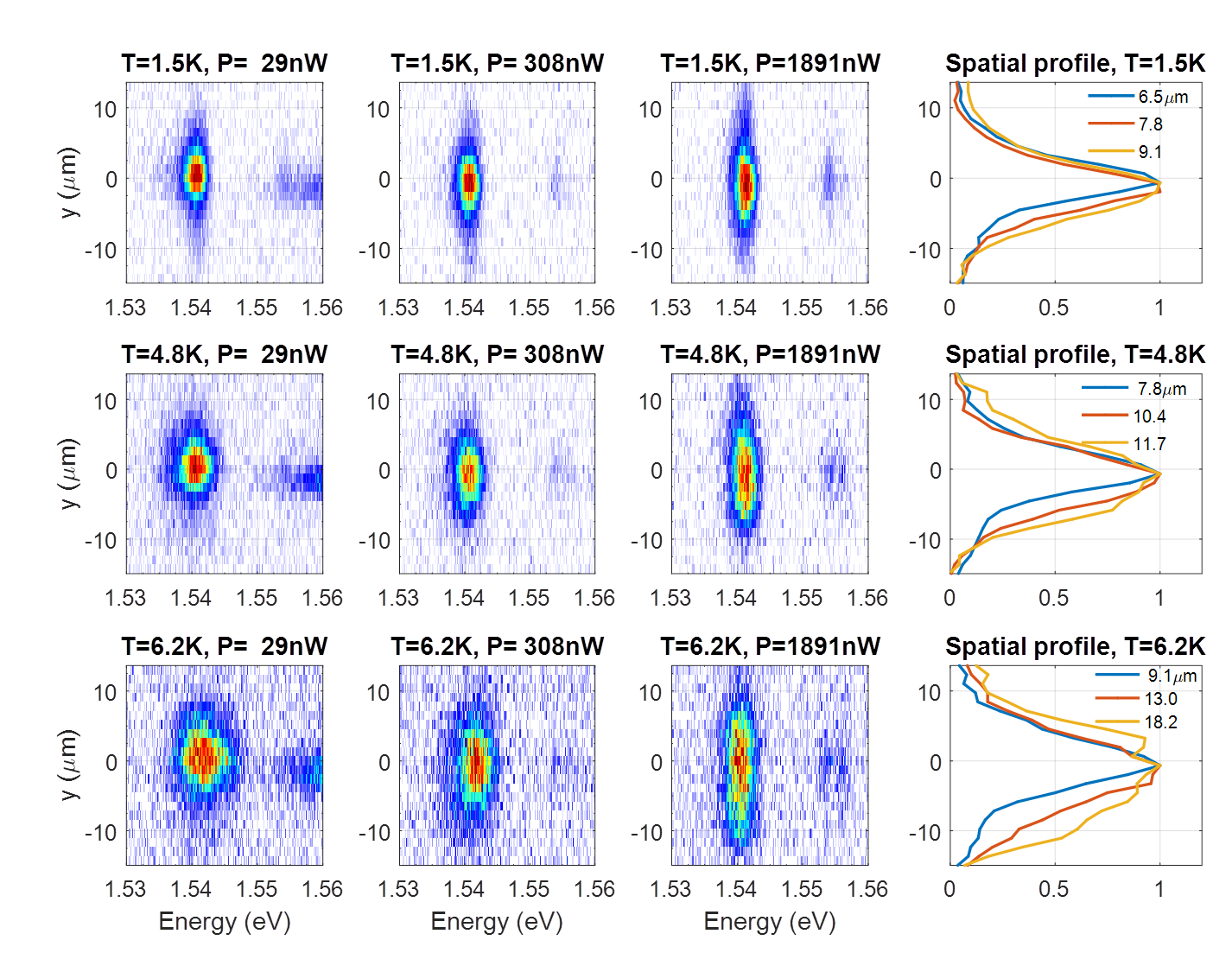}\\
  \caption{Spatial-spectral photoluminescence images recorded for temperatures $T\approx 1.5, 4.8, 6.2K$ and excitation powers $P\approx 29, 308, 1891nW$, all for constant energy line $E_{IX}\approx 1.541eV$. The column to the right shows the corresponding spatial profiles for each temperature, obtained by integration along the spectral axis, and the legend shows FWHM of the curves (color ordering is for low, medium and high power).}
  \label{fig:raw_data_extended_profiles}
\end{figure}

Figure \ref{fig:raw_data_extended_profiles} shows some raw data of photoluminescence, obtained at different temperatures and excitation powers, all for a constant energy line $E_{IX}\approx 1.541eV$. One can see a very minor increase along the Xtrap spatial cross-section for the highest excitation power at $T=1.5K$, which translates into a slightly increased trap area by a factor of less than 1.4.

\section*{Supplementary Note 2 - Extraction of the blue-shift energy $\Delta E$ from the applied field}
\begin{figure}[ht!]
  \centering
  \includegraphics[width=0.5\textwidth]{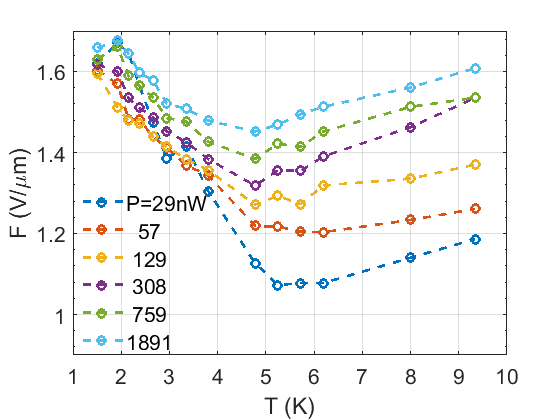}\\
  \caption{The electric field $F$ that is needed in order to maintain a specific constant energy line of $E_{IX}\approx1.541eV$ as a function of temperature, for several different excitation powers.}
  \label{fig:FTallP}
\end{figure}

The electric field $F$ that is needed in order to maintain a specific constant energy line (i.e. a fixed $E_{IX}$) is calculated using $F=V/L$, where $V$ is the applied voltage and $L$ is the active sample thickness. Such data is shown in Fig. \ref{fig:FTallP} as a function of temperatures for several different excitation powers.

\begin{figure}[ht!]
  \centering
  \includegraphics[width=0.5\textwidth]{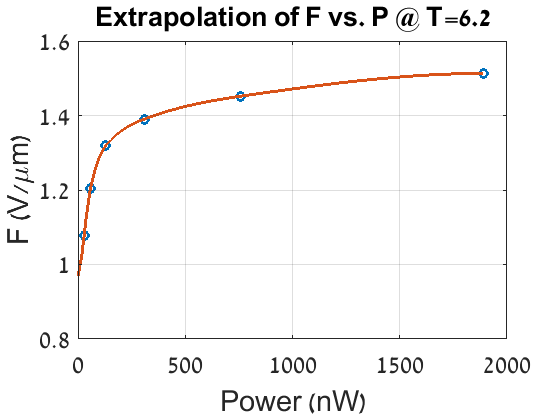}\\
  \caption{The electric field $F$ as a function of excitation power $P$, recorded at a temperature $T\approx 6.2K$. Blue circles indicate measured data points while the red curve is an interpolation (between points) or extrapolation (towards $P=0$) of the data.}
  \label{fig:FPextrap}
\end{figure}
In order to extract the blue shift energy $\Delta E(T,P)$ from $F(T,P)$, we note that
\begin{equation}\label{eq:Eix_TP}
  E_{IX} = E_{DX} - eF(T,P)d + \Delta E(T,P).
\end{equation}
where the direct exciton energy $E_{DX}$ has a very weak dependent on the density, temperature and the applied field (quadratic stark shift) in the measured parameters range.
Now, in the limit of a very weak excitation power the indirect exciton gas in the trap  becomes very dilute, and the blue shift energy is negligible: $\Delta E \rightarrow 0$ so one gets
\begin{equation}\label{eq:Eix_T0}
  E_{IX} = E_{DX} - eF(T,0)d
\end{equation}
For a fixed $E_{DX}-E_{IX}$ (i.e. on a given CEL) we can combine the above equations and get that
\begin{equation}\label{eq:DeltaE_TP}
  \Delta E(T,P)=eF(T,P)d-eF(T,0)d.
\end{equation}
Next, we extrapolate the data of electric field $F$ vs.\ excitation power $P$ towards zero to extract $F(T,0)$ in the limit of zero excitation power, as is shown in Fig. \ref{fig:FPextrap}.

The temperature dependence of $F(T,0)$ is expected to be negligible since the band gap dependence with $T$ is essentially flat in this temperature range and there are no multi-particle interactions in this limiting low density. Therefore we chose the data of $T\approx 6.2K$ for the extrapolation process and the same electric field $F(T=6.2K,0)$ is used to deduce the blue shift energy for all temperatures. This temperature is chosen since for $(T\ge T_c)$ there is a clear power dependence of $F$, and no accumulation of dark particles is expected. The extrapolation of $F(T,P)$ towards $F(T,0)$ for other temperatures above $T\ge 4.8K$ gives similar results within less than $5\%$ error.

\section*{Supplementary Note 3 - Intensity and blue-shift energy for different constant energy lines (CELs)}
\begin{figure}[ht!]
  \centering
  \includegraphics[width=0.5\textwidth]{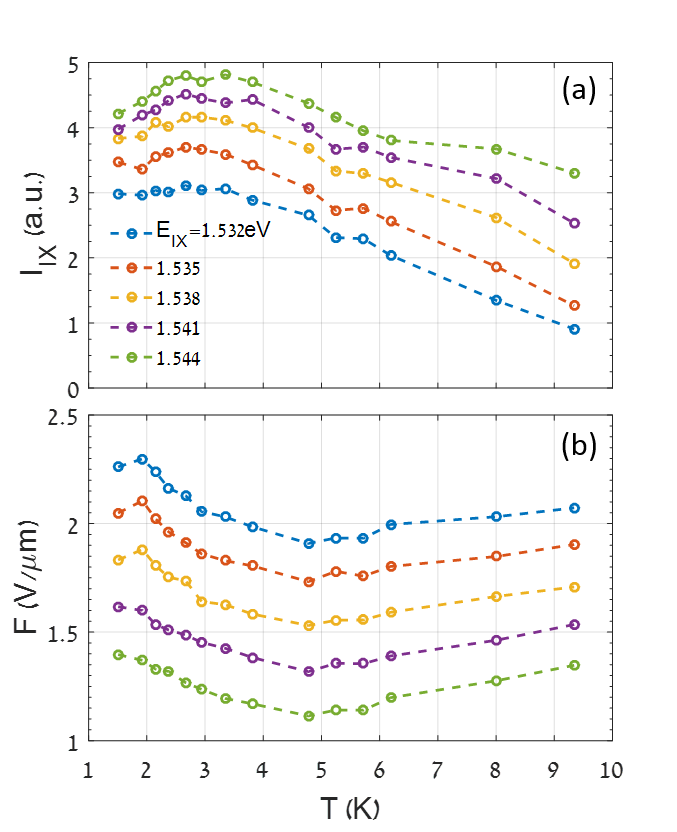}\\
  \caption{(a) Measured PL intensity as a function of temperature for different CEL. All curves are for a fixed excitation power $P=308nW$. (b) The applied electric field $F$ needed in order to maintain the CEL as a function of temperature, for the same set of measurements as in (a).}
  \label{fig:IFT_CEL}
\end{figure}

The results presented in the paper are not restricted to a specific choice of indirect exciton energy $E_{IX}$, but rather show similar features for the whole range of energies that was measured. Figure \ref{fig:IFT_CEL}(a) shows the recorded PL intensity as a function of temperature for different $E_{IX}$ values (at fixed excitation power). Figure \ref{fig:IFT_CEL}(b) shows the applied electric field $F$ needed to maintain the CEL as a function of temperature. As one can expect, lower $E_{IX}$ requires stronger electric field in agreement with the linear Stark effect. 
We can then extract the blue-shift energy $\Delta E$ as is described in Supplementary Note 2. These results are shown in Figure \ref{fig:DeltaE_peacock}, and display essentially the same features as shown in the main text.
\begin{figure}[ht!]
  \centering
  \includegraphics[width=0.8\textwidth]{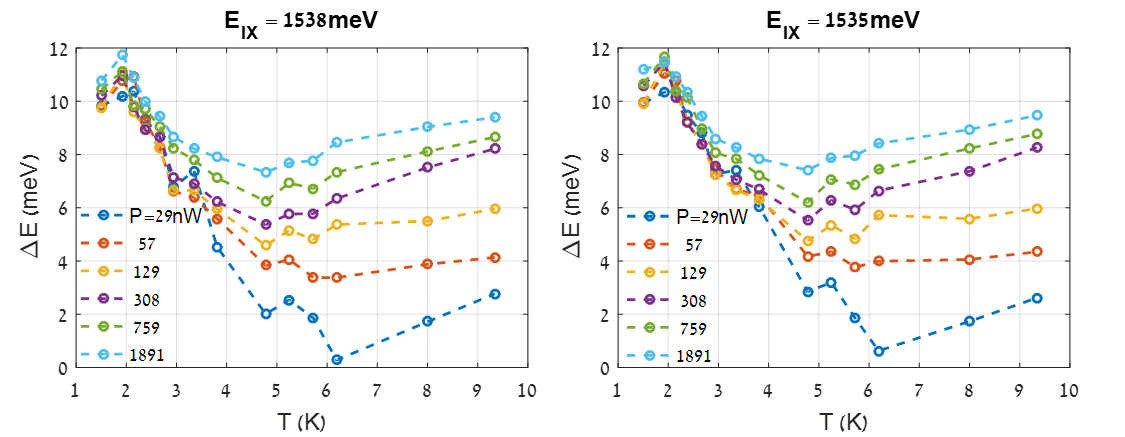}\\
  \caption{The blue shift energy $\Delta E$ as a function of temperature, for several different excitation powers. Results are shown for two CELs with $E_{IX}=1.535eV$ and $E_{IX}=1.538eV$.}
  \label{fig:DeltaE_peacock}
\end{figure}

\section*{Supplementary Note 4 - Photoluminescence intensity vs. tunneling current}
\begin{figure}[ht!]
  \centering
  \includegraphics[width=0.5\textwidth]{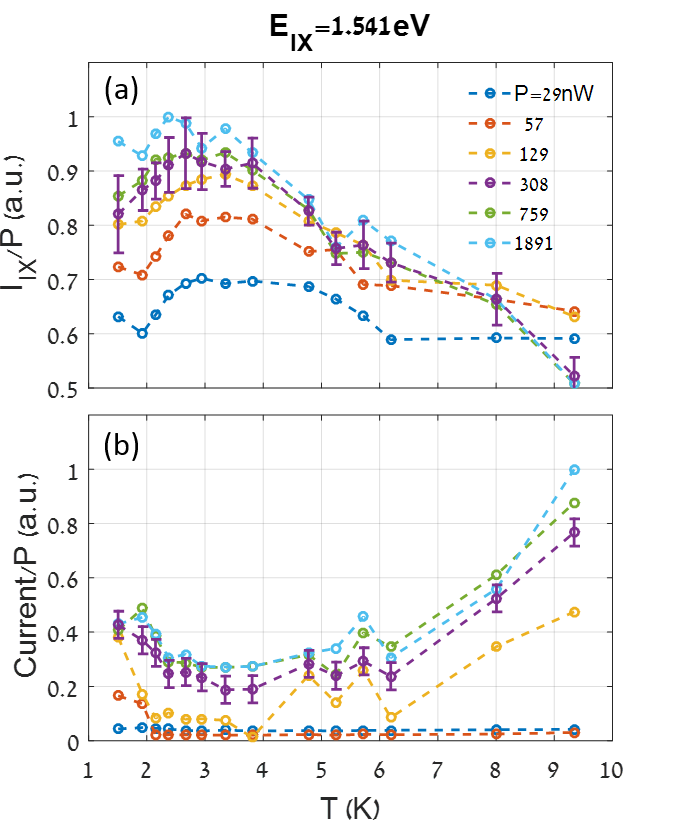}\\
  \caption{(a) The total IX intensity $I_{IX}$ and (b) the DC current flowing through the sample, as a function of temperature for several different excitation powers. The curves in are normalized to their corresponding laser power $P$. Error bars are shown only for a single power level.}
  \label{fig:I_cur_vs_T}
\end{figure}

While maintaining a constant energy line of $E_{IX}=1.541eV$, we recorded the total IX photoluminescence (PL) and tunneling current as a function of temperature for different excitation powers. The results are shown in Fig. \ref{fig:I_cur_vs_T}, where we plot the PL intensity and tunneling current normalized by the excitation power. One can see an increasing PL intensity as the temperature is lowered down to about $T\approx3K$, where the intensity begins to decrease (darken). The tunneling current shows roughly an opposite pattern (except for the two lowest excitation powers, where the actual values are within or below our measurement resolution).
These findings can be explained in light of the competing nature of radiative and non-radiative decay mechanisms. At high temperatures, there is a lower fraction of excitons occupying the light cone, and thus non-radiative mechanisms such as tunneling current are more dominant. As we cool down, more excitons enter the light cone and become radiative, so the intensity increases and the current decreases.

\begin{figure}[ht!]
  \centering
  \includegraphics[width=0.5\textwidth]{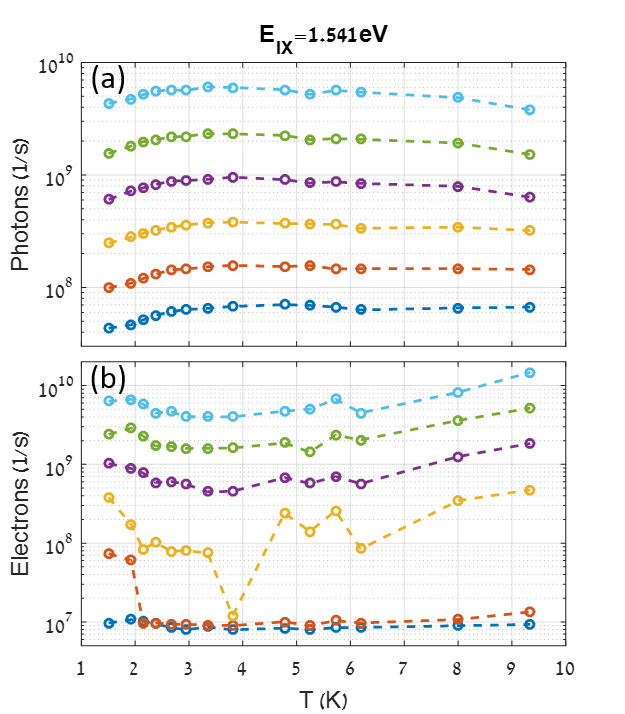}\\
  \caption{(a) The total emitted photons rate and (b) the charge carriers flowing through the sample rate , as a function of temperature for several different excitation powers. Color code is the same as in Fig. \ref{fig:I_cur_vs_T}.}
  \label{fig:photons_electrons_vs_T}
\end{figure}

However, there is a clear change of trend below $T\approx 3K$, from increasing to decreasing intensity as we lower the temperature. This suggests that radiative processes become less probable, as more excitons reside in dark states, and are thus compensated by an increasing tunneling current. To check this hypothesis, we carried out a careful and detailed analysis of the data which allowed us to compare the total number of emitted photons to that of electrons tunneling through the sample (for details see supplementary note 5 next). The results are presented in Fig. \ref{fig:photons_electrons_vs_T} and indeed show a very good agreement between the decrease of the photons' rate at the lower temperature range and the increase of charge carriers rate.
This indicates that the "loss" of photons is accounted for by the increase of e-h pairs in the current (each exciton contributes one e-h pair if it tunnels out non-radiatively), so conservation of particles is maintained.

\section*{Supplementary Note 5 - Bright excitons density estimation}
In order to calculate the bright exciton density $n_b$ from the measured photoluminescence, we assume a thermal distribution of excitons with uniform density across the trap area. In this case, we can write a rate equation for $n_b$ of the following form:
\begin{equation}\label{eq:density_rate}
\frac{\partial n_b}{\partial t} = -\frac{\beta n_b}{\tau} - cn_b^2 + G 
\end{equation}
where $\beta = n_{rad}/n_b$ is the fraction of bright excitons occupying the light cone ($\beta$ is temperature-dependent) \cite{shilo_particle_2013}, $\tau$ is the single exciton radiative lifetime, $G$ is exciton generation rate and $c$ is a bi-molecular recombination coefficient describing radiative emission due to two-particles processes (where $cn_b \ll \beta/\tau$). In addition, due to the limited numerical aperture of our experimental apparatus, only a relatively small fraction of photons that are emitted from recombining bright excitons are collected and reach the detector (CCD). We denote this fraction by $\gamma = n_{collected}/n_b$. 

\begin{figure}[ht!]
  \centering
  \includegraphics[width=0.75\textwidth]{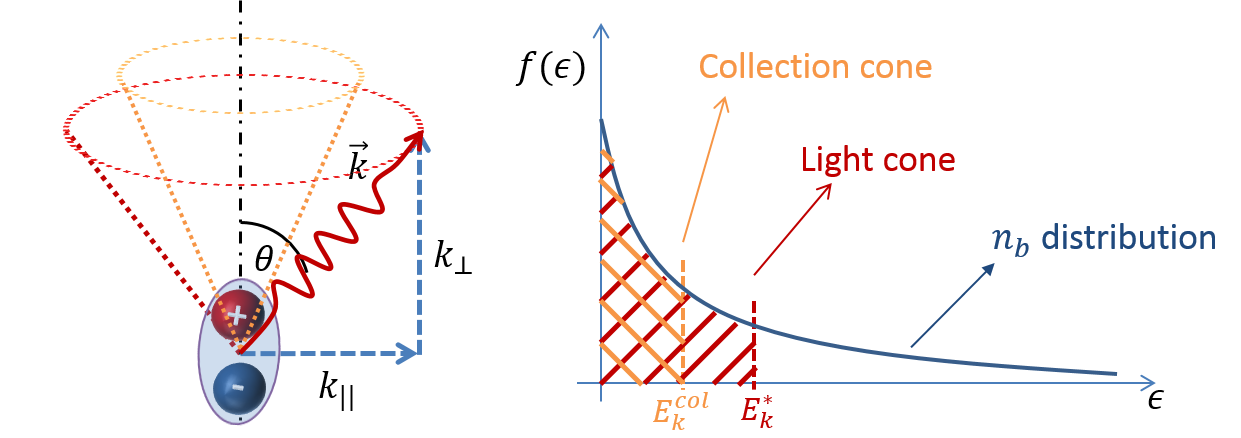}
  \caption{Left: an illustration of a recombining exciton. $k_\parallel$ is the IX's in-plane momentum, while $k_\perp$ and $\vec{k}$ are the photon's perpendicular and total momentum, respectively.
Right: schematics of a thermal distribution of bright IXs kinetic energy. Excitons with energy below some value $E^*$ are inside the light cone and thus radiative, while photons originating from excitons with energy below $E^{col}$ are collected into the numerical aperture of our system.}
  \label{fig:light_cone}
\end{figure}

\begin{figure}[ht!]
  \centering
  \includegraphics[width=0.5\textwidth]{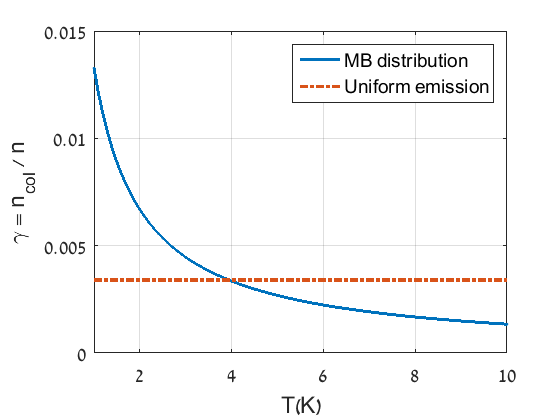}
  \caption{The fraction of excitons whose emitted light is collected by the experimental system numerical aperture of NA=0.42, $\gamma = n_{col}/n_b$, assuming a Maxwell-Boltzmann thermal distribution of excitons with momentum conservation (solid, blue), compared with a uniform emission pattern into a $4\pi$ solid angle (dashed, red).}
  \label{fig:gamma}
\end{figure}

In general, these ratios depend on the distribution of excitons as a function of energy since they arise from conservation of energy and in-plane momentum (see Fig. \ref{fig:light_cone}).
Figure \ref{fig:gamma} shows for example $\gamma$ for the case of a uniform emission distribution (which means no in-plane momentum conservation) compared with a thermal Maxwell-Boltzmann distribution of excitons with momentum conservation. The error bars in the following estimations and those presented in Fig. 4(c) of the main text account for both limiting cases.

\begin{figure}[ht!]
  \centering
  \includegraphics[width=0.5\textwidth]{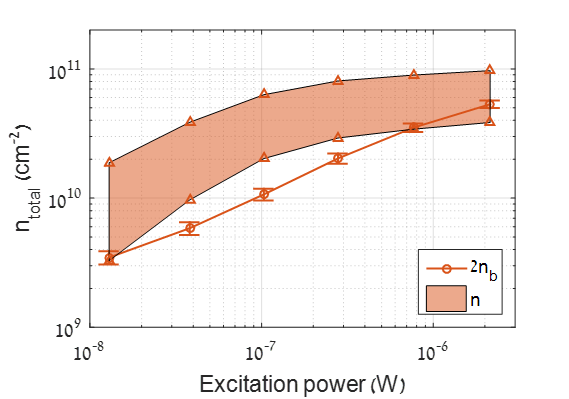}
  \caption{Estimated density of the exciton cloud $n_{total}$ as a function of excitation power at $T=5.4K$ and $E_{IX}=1.538eV$.
Twice the bright exciton density, $2n_b$, obtained from photon-counting (circles) is presented together with the density range (triangles, shaded region) that is calculated from the measured blueshift energy $\Delta E$ limited by an uncorrelated mean-field interactions (lower limit) and the highest correlated state of an excitonic crystal (upper limit).}
  \label{fig:n_total_highT}
\end{figure}

The intensity $I$ (counts/unit time) recorded by our detector is given by
\begin{align}\label{eq:intensity}
I & = \eta \cdot A\left(\frac{\gamma n_b}{\tau} + cn_b^2 \right)
\end{align}
where $A$ is the exciton cloud area which is imaged onto the CCD, and $\eta$ is the combined loss of our experimental setup and the conversion efficiency of photons into counts. After measuring and accounting for all losses incurred by the optical elements of our setup and carefully calibrating the photons/count ratio in our system, we can extract the bright exciton density. $\tau$ was measured independently using a time resolved measurement described in detail in Ref.  \cite{shilo_particle_2013} so the only fitting parameter (i.e. the unknown parameter) in our model is $c$, which is determined by the best agreement between the total density as extracted from $\Delta E$ and that of $2n_b$ for $T=5.4K$, as shown in Fig. \ref{fig:n_total_highT}. We note however that the fit weakly depends on $c$, since assuming $c=0$ only slightly changes the quality of the fit.

\section*{Supplementary Note 6 - Density estimation using an interacting quantum liquid expression}
In the main text, we used the theoretical classical liquid interaction energy adopted from Ref. \cite{laikhtman_exciton_2009} (see Eq. 5.5 of this reference) to deduce the exciton density. However, this expression neglects the contribution from quantum confinement energy that comes from "caging" the particle and thus adding a zero-point energy term which becomes not negligible in this density range. In order to have a more accurate density estimation, we can add the quantum correction to the interaction energy (Eq. 5.2 of Ref. \cite{laikhtman_exciton_2009}) to have:
\begin{equation}\label{eq:quantum_energy}
  \Delta E\approx \underbrace{\frac{10e^2d^2}{\kappa}n^{3/2}}_{E_C}+\underbrace{\sqrt{\frac{30\hbar^2e^2d^2}{\kappa M_X}}n^{5/4}}_{E_Q}
\end{equation}
Solving this equation numerically for a value of $\Delta E=10.5meV$, one gets a density $n\approx9\cdot10^{10}cm^{-2}$. Plugging back this density into the equation yields $E_C=7.7meV$ and $E_Q=2.8meV$, so indeed the kinetic quantum confinement energy is significant.

\section*{Supplementary Note 7 - Spatial coherence measurements}
\begin{figure}[ht!]
  \centering
  \includegraphics[width=0.5\textwidth]{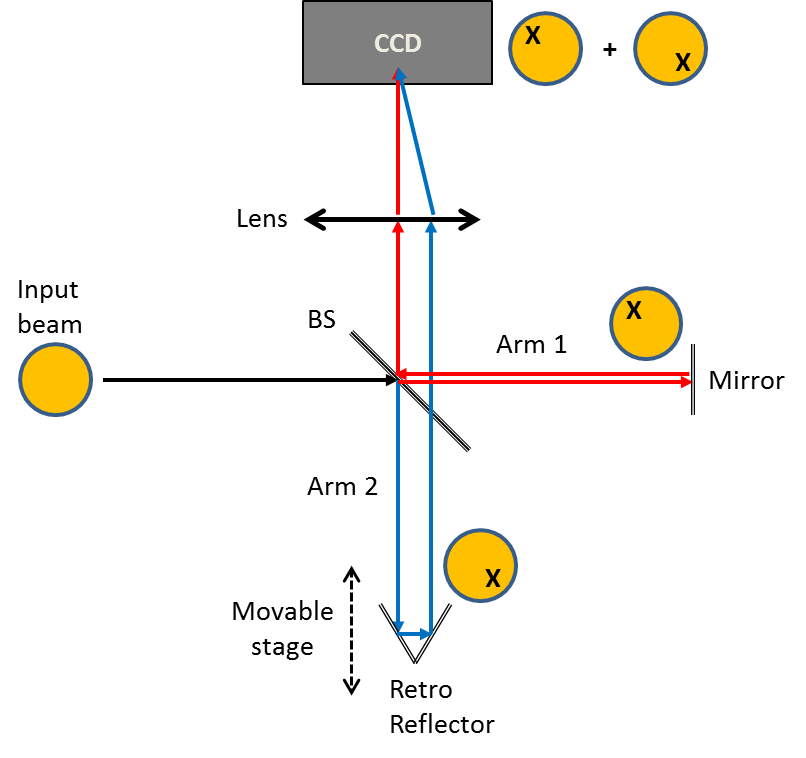}\\
  \caption{A Michelson flip-image interferometer.}
  \label{fig:Michelson}
\end{figure}

We performed spatial coherence measurements on the same IX system discussed in this work, covering a wide range of excitation powers and temperatures, using a flip-image interferometry \cite{kasprzak_bose-einstein_2006}, and were unable to detect any spatial coherence of the PL over distances larger than the spatial resolution of the measurement system ($\sim1\mu m$), even at excitation powers and temperatures where the anomalous behavior reported in this work was found. We used a Michelson flip-image interferometer, shown in Fig. \ref{fig:Michelson}, to measure the spatial coherence of PL emitted from indirect excitons in the trap. The spatial resolution of the optical system as given by $\lambda/2\text{NA}\approx1\mu m$ for our parameters $\lambda\approx800nm$ and $\text{NA}=0.42$. 

The experimental method is as follows:
 \begin{itemize}
   \item Aligning the two arms of the interferometer to a sub-picosecond time delay using an ultra fast laser source (to ensure coherence time longer than the reciprocal of the PL linewidth).
   \item Passing the PL from the sample through this configuration.
   \item Scanning the phase of one arm for few times $2\pi$ while recording images, and detecting the oscillation frequency in each pixel due to the movement of the fringes.
   \item Using Fourier analysis to obtain the amplitude of oscillation in the detected frequency which, after background substraction and proper normalization, represents the degree of coherence.
 \end{itemize}
 
\begin{figure}[ht!]
  \centering
  \includegraphics[width=0.5\textwidth]{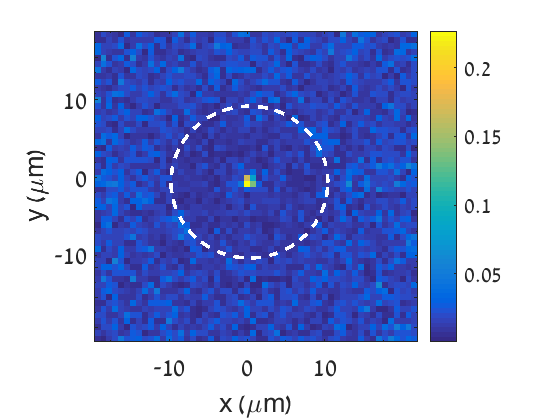}\\
  \caption{A spatial degree of coherence map of the PL from trapped IX. Experimental conditions are $T=1.7K$ and $P=2\mu W$. The dashed line marks the boundaries of the Xtrap.}
  \label{fig:coherence}
\end{figure}

The above procedure was performed over a range of experimental conditions similar to those shown in the main text. Nevertheless, no extended spatial coherence was detected, as illustrated in Fig. \ref{fig:coherence}.
This lack of coherence can be consistent with a formation of a dark liquid, where even if the coherence is not strongly suppressed by the strong interactions, it is mostly manifested in the dark exciton part and therefore no spatial coherence is observed in the bright, emitting part.
We note that other experiment on cold IX fluids were also unable to detect any spatial coherence down to very low fluid temperatures \cite{stern_exciton_2014, repp_confocal_2014}.





\end{document}